\documentclass[aps,prd,nofootinbib,floatfix,twocolumn]{revtex4-2}

\newif\ifonecolumnlayout

\onecolumnlayoutfalse 

\ifonecolumnlayout
  \PassOptionsToClass{onecolumn,notitlepage,11pt}{revtex4-2}
\else
  \PassOptionsToClass{twocolumn,10pt}{revtex4-2}
\fi
\PassOptionsToClass{showkeys,nofootinbib,amsmath,amssymb,aps,pra,floatfix}{revtex4-2}

\ifonecolumnlayout
  \usepackage[a4paper,margin=2.5cm]{geometry}
\fi

\usepackage[toc,page]{appendix}
\usepackage[T1]{fontenc}
\usepackage{newtxtext,newtxmath}
\usepackage{titlesec}
\usepackage{soul}
\usepackage{xcolor}

\usepackage{empheq}
\usepackage{graphicx}
\usepackage{subfig}
\usepackage[english]{babel}
\usepackage{physics}
\usepackage{url}
\usepackage{silence}
\makeatletter
\let\label\ltx@label
\makeatother
\usepackage[colorlinks=true,linkcolor=blue,citecolor=red,urlcolor=blue,filecolor=blue,linktocpage=true]{hyperref}
\hypersetup{allcolors=blue,citecolor=red}
\usepackage{cleveref}
\usepackage[utf8]{inputenc}
\usepackage[toc,page]{appendix}
\usepackage[autostyle]{csquotes}
\usepackage{slashed}
\usepackage{graphicx}
\usepackage{amsmath}
\makeatletter
\let\over\@@over
\makeatother
\usepackage{nccmath}
\usepackage{cancel}
\usepackage{mathtools}
\usepackage{amsfonts}
\usepackage{upgreek}
\usepackage{paralist} 
\DeclareMathAlphabet{\mathpzc}{OT1}{pzc}{m}{it} 
\usepackage{tensor}
\makeatletter
\renewcommand{\@keys@name}{\textsc{Keywords: }}
\makeatother

\titleformat{\section}{\normalfont\large\bfseries\sffamily}{\thesection}{1em}{}
\titleformat{\subsection}{\normalfont\normalsize\bfseries\sffamily}{\thesubsection}{1em}{}
\titleformat{\subsubsection}{\normalfont\normalsize\bfseries\sffamily}{\thesubsubsection}{1em}{}
\titleformat{\paragraph}[runin]{\normalfont\normalsize\bfseries\sffamily}{\theparagraph}{1em}{}[.]
\titlespacing*{\section}{0pt}{2.8ex plus 0.8ex minus .2ex}{1.2ex plus .2ex}
\titlespacing*{\subsection}{0pt}{2.4ex plus 0.8ex minus .2ex}{1.1ex plus .2ex}
\titlespacing*{\subsubsection}{0pt}{2.4ex plus 0.8ex minus .2ex}{1.1ex plus .2ex}
\titlespacing*{\paragraph}{0pt}{1.5ex plus 0.5ex minus .2ex}{1em}
\renewcommand{\thesection}{\arabic{section}}
\renewcommand{\thesubsection}{\thesection.\arabic{subsection}}
\renewcommand{\thesubsubsection}{\thesubsection.\arabic{subsubsection}}

\makeatletter
\def\Dated@name{}

\def\frontmatter@above@affiliation{\par\vspace{10pt}}
\def\frontmatter@above@affilgroup{\par\vspace{10pt}}
\makeatother

\makeatletter
\renewcommand{\@dotsep}{10000}
\renewcommand\l@section[2]{\vskip 0.45em\@dottedtocline{1}{0em}{1.8em}{#1}{{\fontseries{b}\selectfont #2}}}
\renewcommand\l@subsection[2]{\@dottedtocline{2}{1.8em}{2.2em}{#1}{#2}}
\renewcommand\l@subsubsection[2]{\@dottedtocline{3}{4.0em}{2.8em}{#1}{#2}}
\makeatother

\makeatletter
\def\ps@jhepfooter{%
  \let\@oddhead\@empty
  \let\@evenhead\@empty
  \def\@oddfoot{\hfil--\ \thepage\ --\hfil}%
  \def\@evenfoot{\hfil--\ \thepage\ --\hfil}%
}
\makeatother

\begin{document}
\hbadness=10000
\sloppy

\addtolength{\footskip}{6pt}
\pagestyle{jhepfooter}

\ifonecolumnlayout\preprint{}\else\preprint{APS/123-QED}\fi


\title{{\sffamily\bfseries\Large Quantum speed limit time of a uniformly accelerated Unruh-DeWitt detector}}

\author{Debasish Ghosh$^{1}$}
\author{and Bibhas Ranjan Majhi$^{1}$}%
\affiliation{$^{1}$ Department of Physics, Indian Institute of Technology Guwahati, Guwahati 781039, Assam, India}
\affiliation{\vspace{7pt}\textit{E-mail:} {\normalfont\href{mailto:g.debasish@iitg.ac.in}{g.debasish@iitg.ac.in}}, {\normalfont\href{mailto:bibhas.majhi@iitg.ac.in}{bibhas.majhi@iitg.ac.in}}}

\date{}

\begin{abstract}
\noindent Within the open quantum system formalism, using the Markovian process, we investigate the {\it quantum speed limit time} (QSLT) of a uniformly accelerated Unruh-DeWitt detector, interacting with the massless real scaler field. Three specific scenarios are being studied -- the field is in Minkowski spacetime (i) without any reflecting boundary but at a finite temperature, (ii) presence of single boundary with the detector is accelerating parallel to the plane of it, and (iii) presence of two boundaries with the  detector is moving between them along the parallel direction.  Interestingly, the behavior of QSLT appears to be very distinctive around a particular value of $a$. For the first scenario QSLT shows very noticeable change around a specific $a$ and the required value of $a$ corresponding to this change decreases with the increase of field temperature. In the other two scenarios, QSLT starts to fluctuate after a critical value of $a$. If we further increase $a$, more vigorous fluctuation emerges. The value of $a$ required for the starting of the oscillation is much less for the double boundary case compared to the single boundary. Taking into account specific values of the distance between the boundaries and the detector's frequency, we estimate that the critical value of $a$ can be as low as $\sim 10^5 \text{m}/\text{s}^2$ for the double boundary situation. We argue that such a distinctive feature in QSLT can be a fruitful tool to diagnose the Unruh effect.      
\end{abstract}

\maketitle
\ifonecolumnlayout
  \thispagestyle{empty}\setcounter{page}{0}\newpage
\else
  \setcounter{page}{1}\thispagestyle{jhepfooter}
\fi

\begingroup
\hypersetup{linkcolor=blue}
\setlength{\parskip}{0pt}
\endgroup
\hypersetup{linkcolor=blue}


\section{Introduction}\label{Intro}


In quantum mechanics, the energy--time uncertainty relation suffers interpretational difficulties \cite{Deffner:2017cxz} because time is not represented by an operator in the same way as other observables. To address this issue, Mandelstam and Tamm \cite{Mandelstam1991} and later Margolus and Levitin \cite{MARGOLUS1998188} provided an alternative interpretation of the relation. They introduced the concept of the minimum time required for a quantum system to evolve from an initial quantum state to an orthogonal quantum state. This minimum evolution time is known as the {\it quantum speed limit time} (QSLT) (for a review on QSLT, see \cite{Deffner:2017cxz}). Latter, the concept was generalized to mixed states and open quantum systems \cite{Deffner:2017cxz,Campaioli:2019esq,Khan:2020bgi}. QSLT plays important role in various topics in  quantum information theory; e.g. quantum communication, quantum metrology, computational bounds of physical systems, quantum optimal control algorithms, etc. Since then, considerable effort has been devoted in estimating the minimum evolution time of different physical quantum systems \cite{Deffner:2017cxz,Taddei2013,PhysRevLett.110.050403}. In this work, within the open quantum system formalism we investigate the features of QSLT of a uniformly accelerated two-level atom (namely the Unruh-DeWitt detector) moving in Minkowski spacetime and interacting with a background massless real scalar field.

So far investigations have been done to explore the relativistic effects \cite{Villamizar:2015eho} as well as accelerations of the observers' frame \cite{Zhang:2013wfe,Khan:2015aza} on the QSLT of the damped Jaynes-Cummings and the Ohmic-like dephasing like models. The effects of curvature has also been studied \cite{Haseli:2019ktt,Maleki:2019cqn,Xu:2021pzv}. Further, influence of acceleration was noticed for an entangled state of quantum fields \cite{Khan:2020bgi,Xu:2020nhn}. All these illuminates the character of QSLT under relativistic as well as non-inertial motions. Motivated by these facts we are interested to know characteristic features of the QSLT of a uniformly accelerated Unruh-DeWitt detector which is interacting with background scalar fields. The whole theoretical model is based on the open quantum system formalism within the Markovian process \cite{10.1093/acprof:oso/9780199213900.001.0001}, where the field system acts as the environment.      

A uniformly accelerated observer perceives the Minkowski vacuum as a thermal bath with a temperature proportional to its acceleration, a phenomenon known as the Unruh effect \cite{Unruh:1976db}. This is considered as an important prediction of quantum field theory in a non-inertial frame. Further Einstein's equivalence principle connects it to the Hawking effect \cite{Hawking:1975vcx}, providing valuable insights into the quantum properties of black holes. However, the experimental realization of the Unruh effect remains a major challenge because it requires extremely large accelerations. An acceleration of the order $10^{21}\mathrm{m/s^2}$ is needed to produce an Unruh temperature of approximately $1\mathrm{K}$ \cite{Crispino:2007eb}. 

Several attempts have been made, such as use of ultra-intense lasers \cite{PhysRevLett.83.256}, Penning traps \cite{PhysRevLett.61.2113},  optical cavities \cite{PhysRevLett.91.243004,Lochan:2019osm,Stargen:2021vtg}, the supper-radiant burst \cite{Deswal:2025cjw}, etc., to enhance the accelerated induced transition in the atom. Simultaneously, indirect theoretical approaches and corresponding proposals for experiments have also been put forward, such as thermal quivering \cite{PhysRevD.53.7003}, decay of accelerated protons \cite{PhysRevLett.87.151301}, radiation emissions \cite{PhysRevLett.101.110402,PhysRevLett.97.121302}, choice of particular Fock state \cite{PhysRevLett.61.2113}, decoherence in atomic state \cite{Sahota:2026imu}, radiation shift in atomic spectra \cite{Arya:2024qke}, measuring the Pancharatnam--Berry phase of an accelerated two-level system \cite{PhysRevLett.107.131301,Hu_2012,Ghosh:2024mqy,Barman:2024jpc}, estimation through quantum metrology \cite{Wang:2014uoa}, etc. Despite all these attempts we are still far from the goal. Therefore, continuous investigation is in demand to find a way where the signature of the Unruh effect can be traced at a very low acceleration. We like to take the second school of thought -- an indirect way out. In this vein, here we propose that the theoretical model, based on the concept of QSLT, is likely to trace the signature of the Unruh effect.

Three different scenarios are being analyzed. First, we investigate the QSLT of a uniformly accelerated two-level atom interacting with a background thermal bath of inverse temperature $\beta$. Then QSLT is examined in the presence of reflecting boundaries for the background scalar field at zero temperature. Here the detector is accelerating in parallel to the plane of the boundaries. We divide this study into two cases: (i) a single reflecting boundary and (ii) double reflecting boundaries. In all situations the QSLT is being investigated as a function of the acceleration ($a$) of the detector. We are interested to see whether any distinctive characteristic of QSLT emerges at a relatively low value of $a$.

For the first scenario we observe that for the zero temperature background thermal bath QSLT shows significant change around a particular value of $a$. Further increase in bath temperature results the occurrence of the same at comparatively lower $a$. 
Noticeably, the expression for QSLT bears the exchange symmetry between the inverse background temperature $\beta$ and the inverse Unruh temperature $\beta_U$. This indicates that the Unruh thermal bath behaves in the same way as a real thermal bath, providing further evidence for the thermal nature of the Unruh effect. 
Therefore, the behavior of the QSLT as a function of acceleration at fixed background temperature must be qualitatively similar to its behavior as a function of temperature at fixed acceleration. Hence the observation of any distinctive feature in QSLT in terms of acceleration can be effectively interpreted as the indirect hallmark of the Unruh effect. Such a spirit has been put forwarded in various investigations related to this model (for example see, \cite{Kolekar:2013hra,Kolekar:2013xua,Barman_2021,Ghosh:2024mqy}). Here we carry this idea in latter scenarios as well.

 For the single-boundary case, we place a reflecting boundary at $z=0$ and accelerate the atom parallel to the boundary along the $x$-axis while keeping its distance from the boundary fixed at $z_0$. Interestingly, the presence of the boundary introduces a fluctuating behavior in the QSLT as a function of acceleration. Significant fluctuation starts after a critical value of $a$, and such is more prominent as one increases the acceleration. The observation of this distinguishing effect in QSLT after a critical value of $a$ can be interpreted as an indirect signature of the Unruh effect. Our theoretical prediction indicates that for $z_0=10~\mathrm{km}$, this distinguishing feature can be observable at an acceleration of about $10^{11}\mathrm{m/s^2}$. Furthermore, the required acceleration can be reduced by increasing the distance $z_0$ between the atom and the reflecting boundary. For the double-boundary case, we place two reflecting boundaries at $z=0$ and $z=L$. The atom is accelerated parallel to the boundaries along the $x$-axis while remaining at a fixed distance $z_0$ from the first boundary. Now the significant oscillation can be observed at a relatively small value of $a$. The theoretical estimation shows that for $z_0=30~\mathrm{m}$ and $L=300~\mathrm{m}$, a significant oscillation in the QSLT starts at the low-acceleration region (around $a\sim 10^{5} \mathrm{m/s^2}$). This investigation indicates a possibility of experimental observations of these features through QSLT at a significantly low acceleration. At the end we provide a sketch to measure QSLT. 

 The organization of the paper is as follows. In the next section, we introduce the Unruh-DeWitt model to be investigated. Along with this we find the general expression of the QSLT for this model. Section \ref{Free} is devoted for the analysis of the QSLT of the detector which is in thermal bath. Section \ref{Refl} discusses the effects of the reflecting boundaries on the QSLT. Also a possible way to measure the QSLT is described in Section \ref{Exp}. Finally, we conclude in Section \ref{Concl}. In support of our claims and for few details of the calculation, five appendices are being included at the end.

\section{The model and the quantum speed limit time}\label{Model} 
We first describe the model setup under study. Since this has been studied extensively in literature, a very brief description will be presented. However, the necessary results will be mentioned below which serve our main purpose. Next we calculate QSLT for this system. 

\subsection{The setup}\label{Model1}
Consider an Unruh--DeWitt detector, modeled as a two-level atom, which is interacting with the background massless real scalar field. The quantum field configuration acts as the quantum environment. If the Hamiltonians of the detector (system, $c$) and the environment ($e$) are denoted by $\hat{H}_{c}$ and $\hat{H}_{e}$, respectively, then the total Hamiltonian of the composite system is given by $\hat{H} = \hat{H}_{c} + \hat{H}_{e} + \hat{H}_{\mathrm{int}}$, where $\hat{H}_{\mathrm{int}}$ is the interaction Hamiltonian. The system Hamiltonian is taken as $\hat{H}_{c} = \frac{1}{2}\hbar \omega_{0}\hat{\sigma}_{3}$, where $\hbar \omega_{0}$ is the energy difference between the two levels of the atom and $\hat{\sigma}_{3}$ is the third Pauli matrix. The interaction between the system ($c$) and the environment ($e$) is considered to be $\hat{H}_{\mathrm{int}} = g \phi(x)\hat{\sigma}_{2}$, where $g$ is the coupling constant and $\hat{\sigma}_{2}$ is the operator representing the detector. Similar model has also been chosen in \cite{Ghosh:2024mqy,Barman:2024jpc}. The operator $\hat{\sigma}_{2}$ plays a role analogous to the dipole operator in electromagnetic interactions \cite{Barman:2024jpc}.

We consider initially the environment was in the vacuum state and system was in the state $\ket{\psi(0)}$. The evolution of the density matrix $\hat{\rho}_{ce}$ for the total system--environment composite system is determined by
\[
\frac{d}{dt} \hat{\rho}_{ce}(t) = -i \left[ \hat{H}_{\mathrm{int}}(t), \hat{\rho}_{ce}(t) \right].
\]
However, we are interested only in the dynamics of our system. Therefore, to obtain the reduced dynamics of the system, we trace out the background environment. Within the Markovian process, application of the Born and rotating wave approximations yields the Kossakowski--Lindblad equation \cite{10.1093/acprof:oso/9780199213900.001.0001,Gorini1976CompletelyPD,Benatti2003,Ghosh:2024mqy, Barman:2024jpc,Hu_2012}. In Schrodinger picture this is given by
\begin{equation} \label{eq1}
     \dfrac{d}{dt} \hat{\rho}_c(t) =  - i [\hat{H}_{\text{eff}}, \hat{\rho}_c(t)] + \hat{D}(\hat{\rho}_c(t))~,
\end{equation}
where $\hat{\rho}_c$ is the reduced density operator of the system, and $\hat{H}_{\mathrm{eff}} = \hat{H}_c + \hat{H}_{\mathrm{LS}}$ is the sum of the system Hamiltonian and the Lamb shift Hamiltonian. The Hamiltonian $\hat{H}_{\mathrm{LS}}$ leads to the renormalization of the unperturbed system Hamiltonian. The term $\hat{D}(\hat{\rho}_c(t))$ is known as the dissipation term. 
For the present choice of Hamiltonian the Lamb shifted Hamiltonian is $\hat{H}_{\mathrm{LS}} = \frac{1}{2}\hbar \omega_L \hat{\sigma}_3$ with $\omega_L$ is the Lamb shift factor. So the effective Hamiltonian is $\frac{1}{2} \hbar \Omega \hat{\sigma}_3$, where $\Omega = \omega_0 +\omega_L$. Usually, $\omega_L \ll \omega_0$ so we will neglect $\omega_L$ in further analysis. For our two level system, the dissipation term takes the form
\begin{equation}
     \hat{ D}[\hat{\rho}_c(t)] = \dfrac{1}{2} \sum_{i,j = 1}^{3} a_{ij} 
    \left( 2 \hat{\sigma}_j \hat{\rho}_c \hat{\sigma}_i - \hat{\sigma}_i \hat{\sigma}_j \hat{\rho}_c - \hat{\rho}_c \hat{\sigma}_i \hat{\sigma}_j \right).
\end{equation}
Here $a_{ij}$ is called as component of Kossakowski matrix, given by
\begin{equation} \label{eq3}
    a_{ij} = A \delta_{ij} - iB \epsilon_{ijk} \delta_{k3} + C \delta_{i3} \delta_{j3}.
\end{equation}
The coefficients $A$ and $B$ are determined as,
\begin{equation}\label{eq4}
    A = \dfrac{g^2}{2} \left[ \gamma(\omega_0) + \gamma(-\omega_0) \right], \quad
    B = \dfrac{g^2}{2} \left[ \gamma(\omega_0) - \gamma(-\omega_0) \right]~,
\end{equation}
where $\gamma$ is the Fourier transform of the positive frequency Wightman function $G^{+}(x,x') = \bra{0}\phi(t,\vec{x})\phi(t',\vec{x}\,')\ket{0}$ of the field, i.e.
\begin{equation}\label{eq5}
    \gamma(y) = \int_{-\infty}^{\infty} ds \, e^{i y s} 
    \bra{0} \phi(t,x) \phi(t', x') \ket{0}~,
\end{equation}
with, $s = (t - t')$. Note that the two-point correlation function in Eq.~(\ref{eq5}) is time-translation invariant; therefore, $G^{+}(x,x')$ does not depend on the individual initial and final times, but only on their difference. On the other hand, we have $C = -A$.

For the choice of initial state of the detector as 
\begin{equation} \label{eq6}
   \ket{\psi(0)} = \cos\left(\dfrac{\theta}{2}\right) \ket{0} + \sin\left(\dfrac{\theta}{2}\right) \ket{1}~,
\end{equation}
where, $\ket{0}$ and $\ket{1}$ denote the ground and excited states of the detector respectively, the reduced density matrix at latter time $t=\tau$ is determined from Eq.~(\ref{eq1}). The details of this whole analysis can be followed from \cite{Hu_2012,Ghosh:2024mqy,Barman:2024jpc}. The final result turns out to be 
\begin{widetext}
\begin{equation}\label{eq7}
\hat{\rho}_c(\tau)=
\begin{pmatrix}
\cos^2\left(\dfrac{\theta}{2}\right) e^{-4A \tau} 
- \dfrac{(A-B)}{2A} \left( e^{-4A \tau} - 1 \right)
&
\dfrac{1}{2} e^{-2A\tau - i \Omega\tau} \sin\theta \\
\dfrac{1}{2} e^{-2A\tau + i \Omega\tau} \sin\theta
&
1 - \cos^2\left(\dfrac{\theta}{2}\right) e^{-4A \tau}
+ \dfrac{(A-B)}{2A} \left( e^{-4A \tau} - 1 \right)
\end{pmatrix}~,
\end{equation}
\end{widetext}
where $\tau$ is identified as the detector's proper time.

\subsection{The QSLT}\label{Model2}
The concept of QSLT, which represents the minimum time required for a quantum state to evolve into another quantum state, is a very important in quantum mechanics. It determines the maximum speed at which a quantum system can evolve. Several notions have been proposed to determine the QSLT, such as the Mandelstam--Tamm and Margolus--Levitin bounds. However, since we are dealing with an open quantum system, we employ an expression for the QSLT that is suitable for such systems (see \cite{Khan:2020bgi,Deffner:2017cxz} for a review on this subject). 
If initial state of our system is $\hat{\rho}(0)$ and the evolved final mixed state is $\hat{\rho}({\tau})$, then the general expression for the QSLT of an open quantum system can be written as \cite{Campaioli:2019esq},
\begin{equation}\label{eq8}
\tau_{\mathrm{QSLT}}
=
\frac{
\left\|\hat{\rho}(0)-\hat{\rho}(\tau)\right\|_{\mathrm{HS}}
}{
\overline{\left\|\dot{\hat{\rho}}(\tau)\right\|}_{\mathrm{HS}}
}~,
\end{equation}
where
\begin{equation}
\overline{\left\|\dot{\hat{\rho}}(\tau)\right\|}_{\mathrm{HS}}
=
\frac{1}{\tau}
\int_{0}^{\tau}
\left\|\dot{\hat{\rho}}(\tau')\right\|_{\mathrm{HS}}\,d\tau'.
\label{9}
\end{equation}
The notation $\|\hat{A}\|_{\mathrm{HS}}$ denotes the Hilbert--Schmidt norm of the operator $\hat{A}$, which is defined as $\|\hat{A}\|_{\mathrm{HS}}
=
\sqrt{\mathrm{Tr}\!\left(\hat{A}^{\dagger}\hat{A}\right)}$. The numerator in Eq. (\ref{eq8}) represents the Euclidean distance between the initial and final states, while the denominator corresponds to the average speed of the quantum evolution over the time interval considered.

Substituting equation (\ref{eq6}) and (\ref{eq7}) into equation (\ref{eq8}), we find the expression for QSLT for this model as 
\begin{widetext}
\begin{equation}\label{eq10}
\frac{\tau_{\mathrm{QSLT}}}{\tau}
=
-\frac{
2A
\sqrt{
2\left(e^{-4A\tau}-1\right)^2 F^2
+\frac{1}{2}\left(1+e^{-4A\tau}-2e^{-2A\tau}\cos(\Omega\tau)\right)\sin^2\theta}
}{
D(\tau)
}~.
\end{equation}
In the above $F$ is given by 
\begin{equation}
F=\frac{B-A}{2A}+\cos^2\frac{\theta}{2}~,
\label{11}
\end{equation}
and the denominator turns out to be
\begin{equation}
\begin{aligned}
D(\tau)=\;&
\frac{e^{-2A\tau}}{2}
\sqrt{
\frac{1}{2}(4A^2+\Omega^2) \sin^2\theta
+32A^2F^2e^{-4A\tau}
}
-\frac{1}{2}
\sqrt{
\frac{1}{2}(4A^2+\Omega^2) \sin^2\theta
+32A^2F^2
}
\\[4pt]
&+
\frac{
\frac{1}{2}(4A^2+\Omega^2)\sin^2\theta
}{
8\sqrt{2}AF
}
\ln\left|
\frac{
4\sqrt{2}AF\,e^{-2A\tau}
+
\sqrt{
\frac{1}{2}(4A^2+\Omega^2)\sin^2\theta
+32A^2F^2e^{-4A\tau}
}
}{
4\sqrt{2}AF
+
\sqrt{
\frac{1}{2}(4A^2+\Omega^2)\sin^2\theta
+32A^2F^2
}
}
\right|~.
\end{aligned}
\end{equation}
\end{widetext}
Equation~(\ref{eq10}) is our working expression. Using Eq.~(\ref{eq4}) we will determine the coefficients $A$ and $B$ to demonstrate the behavior of QSLT for different scenarios. As usually happens $\omega_L \ll \omega_0$, in the subsequent discussion $\Omega \simeq \omega_0$ will be considered. A detailed calculation to achieve the working expression is presented in Appendix \ref{App1}.

As mentioned in the introduction, a direct detection of the Unruh effect is almost impossible within the current technology as it requires a very high acceleration ($\sim 10^{21}$ m/s$^2$). Therefore, finding indirect ways of doing so is a growing area. Here we are interested in realizing such a possibility through the detection of QSLT.
 Therefore, we will analyze the QSLT for our system in different scenarios when the atom is uniformly accelerating on the Minkowski spacetime. In particular, we are interested in investigating whether any distinctive feature of QSLT emerges at comparatively low acceleration. Here we target two specific situations. In one case the spacetime is devoid of any reflecting boundary (commonly called as mirror, which can work as a reflecting wall for the field modes). In this case we will discuss the influence of field temperature as well. This will motivate how to realize Unruh efect through QSLT. Other situation will be the influence of mirror (single as well as double), kept perpendicular to the direction of motion of the atom, on the QSLT.

\section{Free spacetime}\label{Free}
In this section, we investigate the QSLT for an accelerated observer interacting with a thermal bath in absence of any mirror. 

\subsection{Analytical expression}\label{Free1}
For the fields are in thermal equilibrium with a bath of inverse temperature parameter $\beta = \frac{1}{k_B T}$, where $T$ is the temperature of the thermal bath, the thermal Wightman function is determined by
\begin{equation}
G^+_{\beta} (x_2,x_1) = \frac{1}{z} \text{Tr} \left\{ e^{-\beta \hat{H}_e} \phi(x_2) \phi(x_1) \right\}.
\end{equation}
Here $z = \text{Tr} \{ e^{-\beta \hat{H}_e} \}$ is the partition function. The trajectory of the atom is given by the Rindler coordinate transformations
\begin{equation}
t = \frac{e^{a\xi}}{a} \sinh(a\eta), 
\qquad
x = \frac{e^{a\xi}}{a} \cosh(a\eta).
\label{eq14}
\end{equation}
$(t,x)$ are the Minkowski coordinates while $(\eta,\xi)$ are the Rindler ones. In order to assure that the accelerated atom is also in thermal equilibrium, the Wightman function must be time translational invariant with respect to the proper time of the Rindler frame. In this case Unruh mode decomposition of the scalar field is convenient (see \cite{Barman_2021,Barman:2021bbw, Ghosh:2024mqy} for a detailed discussion on this context). 
\begin{widetext}
The positive-frequency thermal Wightman function in $(3+1)$ spacetime dimensions is obtained as \cite{Barman_2021}
\begin{equation}\label{eq16}
G^+_{\beta R}
=
\int_0^{\infty} d\omega \int \frac{d^2k_p}{(2\pi)^4}\frac{2}{a}
\left[
\frac{e^{-i\omega\Delta\eta}e^{\frac{\pi\omega}{a}}+e^{i\omega\Delta\eta}e^{-\frac{\pi\omega}{a}}}{1-e^{-\beta\omega}}
+
\frac{e^{i\omega\Delta\eta}e^{\frac{\pi\omega}{a}}+e^{-i\omega\Delta\eta}e^{-\frac{\pi\omega}{a}}}{e^{\beta\omega}-1}
\right]
\kappa\!\left(\frac{i\omega}{a},\frac{k_p e^{a\xi}}{a}\right)
\kappa\!\left(\frac{i\omega}{a},\frac{k_p e^{a\xi}}{a}\right).
\end{equation}
Here $\eta$ is identified as proper time $\tau$. $\kappa[n,z]$ is the modified Basel function with second kind of order $n$.
\end{widetext}
Substituting Eq. (\ref{eq16}) into the expression (\ref{eq5}) one finds, 
\begin{equation}\label{eq17}
    \begin{split}
        \gamma(\omega_0) = \dfrac{\omega_0}{2 \pi}
\dfrac{
\left[
\dfrac{e^{\frac{\pi \omega_0}{a}}}{1 - e^{-\beta \omega_0}}
+ \dfrac{e^{-\frac{\pi \omega_0}{a}}}{e^{\beta \omega_0} - 1}
\right]
}{
2 \sinh\left(\dfrac{\pi \omega_0}{a}\right)
}~;
    \end{split}
\end{equation} 
and 
\begin{equation}\label{eq18}
\begin{split}
\gamma(-\omega_0) =
\dfrac{\omega_0}{2 \pi}
\dfrac{
\left[
\dfrac{e^{-\frac{\pi \omega_0}{a}}}{1 - e^{-\beta \omega_0}}
+ \dfrac{e^{\frac{\pi \omega_0}{a}}}{e^{\beta \omega_0} - 1}
\right]
}{
2 \sinh\left(\dfrac{\pi \omega_0}{a}\right)
}~.
\end{split}
\end{equation}
Then upon substituting Eqs.~(\ref{eq17}) and (\ref{eq18}) into Eq.~(\ref{eq4}), we obtain
\begin{equation}\label{eq19}
   A = \chi_0 \,
\coth\left(\dfrac{\pi \omega_0}{a}\right)
\coth\left(\dfrac{\beta \omega_0}{2}\right),\,\,\,\
  B= \chi_0~;
\end{equation}
where $\chi_0 = \dfrac{g^2 \omega_0}{4 \pi}$ (see also \cite{Ghosh:2024mqy}). Now, by substituting Eq.~(\ref{eq19}) into Eq.~(\ref{eq10}), we can obtain the analytic expression for the QSLT.

Note that Eq.~(\ref{eq19}) is symmetric under the exchange of the inverse temperature of the thermal bath, $\beta$, and the inverse Unruh temperature, $\beta_{\mathrm{U}} = 2\pi/a$. As a result, the QSLT also bears this exchange symmetry. This suggests that the individual effects of the background temperature and the Unruh temperature on the QSLT are equivalent when other one is kept fixed. Hence the observation of any distinctive feature in QSLT in terms of acceleration can be effectively interpreted as the indirect hallmark of the Unruh effect. This idea has been taken in various investigations related to this model (for example see, \cite{Kolekar:2013aka,Kolekar:2013hra,Ghosh:2024mqy}). Here we will carry this idea in latter scenarios as well to identify the signature of Unruh effect.

\subsection{Numerical analysis}\label{Free2}
We now analyze the obtained expression of QSLT numerically. Before going into this, note that the previously obtained expressions ($\tau_{\mathrm{QSLT}}/\tau$, $A$, and $B$), are expressed in the natural units in which $\hbar = c = 1$. However, for the numerical analysis, these constants are restored and included explicitly. Note that the coefficients $A$ and $B$ must have the dimension of inverse time. Therefore, $\chi_0$ must also have the dimension of inverse time. Since $\omega_0$ already has the dimension of inverse time, the remaining factor in $\chi_0$ must be dimensionless. From the interaction Hamiltonian $\hat{H}_{\mathrm{int}} = g\,\phi(x)\hat{\sigma}_2$, together with the field dimension $\phi \sim \sqrt{\hbar/(\omega V)}$, one finds that $[g^2]=[\hbar c^3]$. Hence, to restore $c$ and $\hbar$ explicitly in the expressions (\ref{eq18}) and (\ref{eq19}), we  need the following replacements: $\chi_0 = \frac{g^2\omega_0}{4\pi}$ $\to$ $\chi_0 = \frac{g^2\omega_0}{4\pi\hbar c^3}$, where the factor $\dfrac{g^2}{4\pi\hbar c^3}$ is dimensionless,  $\frac{\pi \omega_0}{a}$ $\to$ $\frac{\pi \omega_0 c}{a}$ and $\frac{\beta \omega_0}{2}$ $\to$ $\frac{\hbar\omega_0}{2 k_B T}$.   

To plot the function $\tau_{\mathrm{QSLT}}/\tau$, we need to estimate the numerical values of several input parameters. In the following, we describe how these values are obtained. First, we choose the transition frequency of the qubit to be a typical value $\omega_0 \sim 1~\mathrm{GHz}$ (for instance in superconducting flux qubits such a value corresponds to a system energy scale of $H_s \simeq \hbar\omega_0 \sim 10^{-25}~\mathrm{J}$ \cite{Paauw2009}). Since the perturbative treatment is valid only when the interaction Hamiltonian is much smaller than the system Hamiltonian, i.e., $H_I \ll H_s$, we require $H_I \ll 10^{-25}~\mathrm{J}$. This implies $g^2\langle \hat{\phi}' \hat{\phi}\rangle \ll 10^{-50}~\mathrm{J}^2$. Usually one has $\langle \hat{\phi}'\hat{\phi}\rangle \sim \hbar/4\pi^2c \sim 10^{-44}$, and therefore we find $g^2 \ll 10^{-6}$. So the constant $g^2/(\hbar c^3) \ll 10^4$. Consequently, for the numerical calculations, we take $g^2/(\hbar c^3) \simeq 1$ {\footnote{If one includes thermal contribution in $\langle \hat{\phi}' \hat{\phi}\rangle$ as well as it is seen from the accelerated frame, then this will be further suppressed by Bosonic factor. In this case the choice $g^2/(\hbar c^3) \simeq 1$ remains well within the perturbative region. Therefore consideration of this value in the numerical analysis is well suited here.}}. The evolution time $\tau$ can be chosen as follows.
Since the characteristic energy scale of the system is $\Delta E \sim \hbar\omega_0$, the uncertainty relation $\Delta E\,\Delta \tau \gtrsim \hbar$ implies that $\Delta\tau \gtrsim 1/\omega_0$. Therefore, for $\omega_0 \sim 10^9~\mathrm{s^{-1}}$, the evolution time must satisfy $\tau \gtrsim 10^{-9}~\mathrm{s}$. In support of these choices of parameters, see Appendix \ref{App2}, where we show that the input values are consistent with the perturbative approximation.

In all the numerical analysis we choose $\tau = 10^{-9}$s. However as it is the minimum value of the uncertainty in time, this can be interpreted as the analogous of Mandelstam-Tamm quantum speed limit time (MT-QSLT) \cite{Mandelstam1991} of our atomic system. So this is the internal time scale of the detector which is being modified by the quantum nature of the environment and the acceleration and leads to $\tau_{\text{QSLT}}$. Hence the ratio (\ref{eq10}) characteristics the modification of the QSL time as a function of the acceleration of the atom.

\begin{widetext}
First, we study how the QSLT changes with acceleration at different temperature. 
\begin{figure*}[http!]
\centering

\subfloat[ ]{
\includegraphics[width=0.42\textwidth,height=5.5cm]{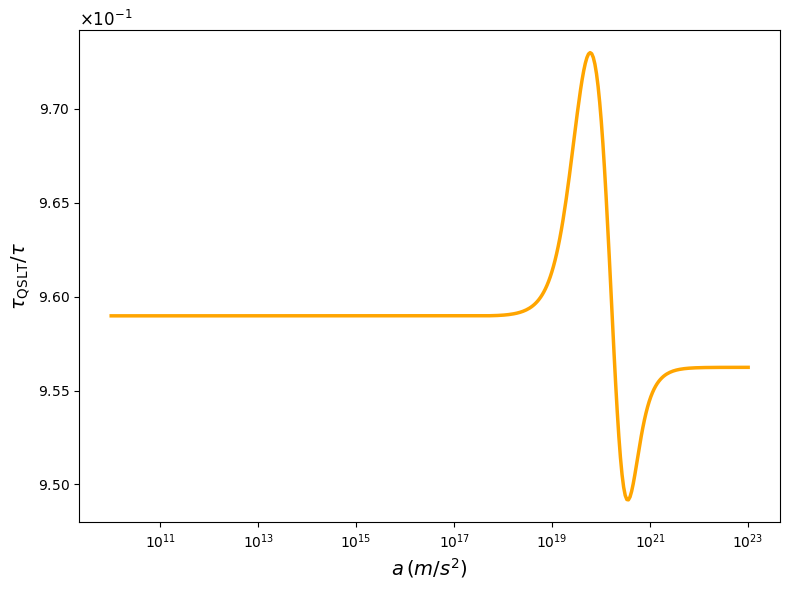}
}
\hfill
\subfloat[ ]{
\includegraphics[width=0.42\textwidth,height=5.5cm]{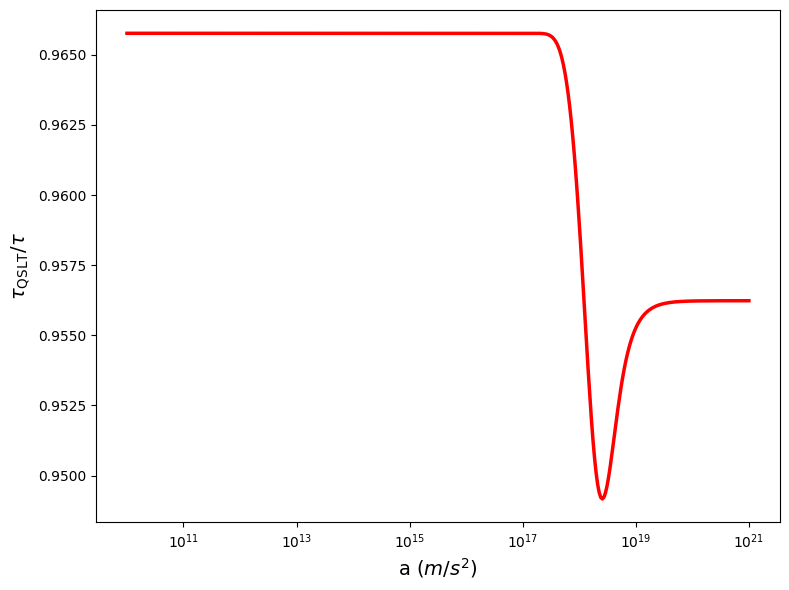}
}

\vspace{0.05cm}

\subfloat[ ]{
\includegraphics[width=0.42\textwidth,height=5.5cm]{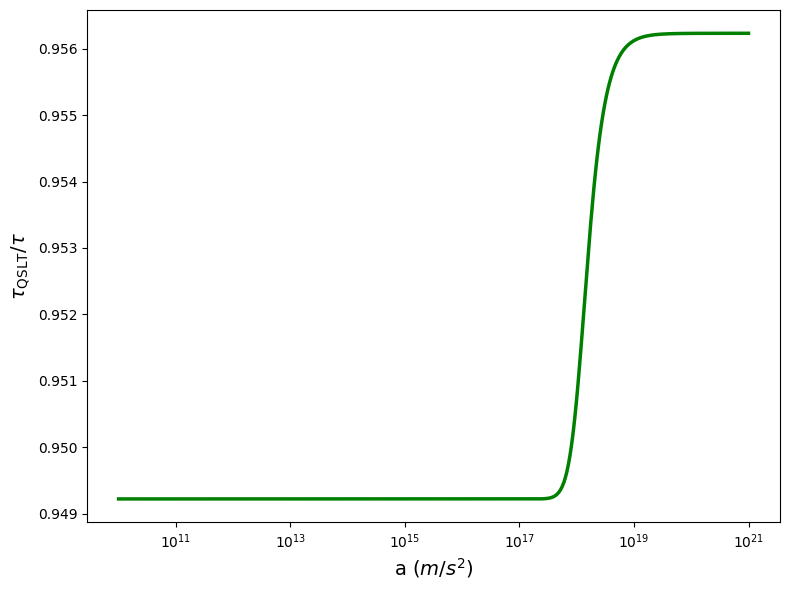}
}
\hfill
\subfloat[ ]{
\includegraphics[width=0.42\textwidth,height=5.5cm]{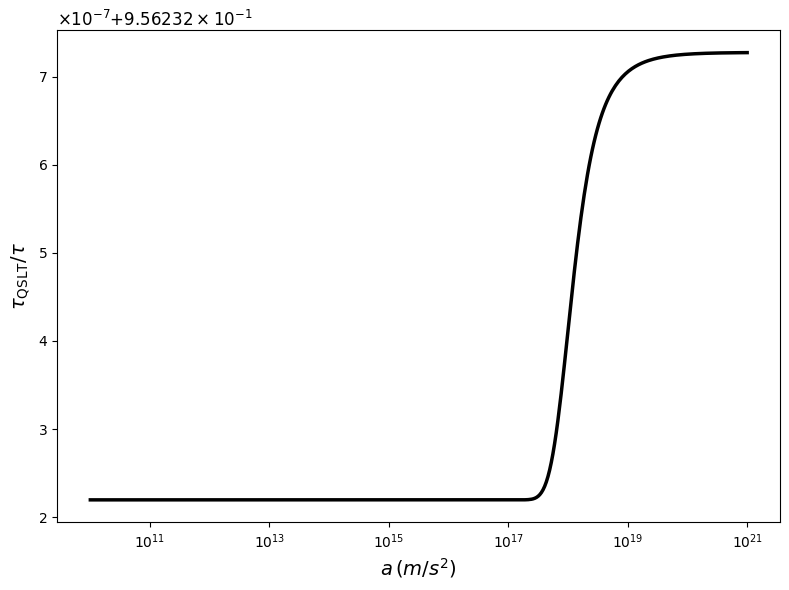}
}

\caption{Variation of the normalized quantum speed limit time, $\tau_{\mathrm{QSLT}}/\tau$, as a function of acceleration for different background temperatures: (a) $0~\mathrm{K}$, (b) $0.5~\mathrm{K}$, (c) $1.5~\mathrm{K}$, and (d) $300~\mathrm{K}$. The remaining parameters are fixed at $\theta=\pi/4$, $\tau=10^{-9}\,\mathrm{s}$, and $g=10^{-5}$.} 
\label{fig:thermalcase}
\end{figure*}
\end{widetext}
In Fig.~\ref{fig:thermalcase}(a), we observe that for a background temperature of $0~\mathrm{K}$, the effects on the QSLT become noticeable around the accelerations of the order of $10^{20}~\mathrm{m/s^2}$. However, when a small background temperature is introduced, as shown in Fig.~\ref{fig:thermalcase}(b), the effect becomes noticeable at lower accelerations, around $10^{18}~\mathrm{m/s^2}$. For a slightly higher temperature, shown in Fig.~\ref{fig:thermalcase}(c), noticeable effects still occur around $10^{18}~\mathrm{m/s^2}$, but the behavior of the QSLT changes. At low temperatures, increasing acceleration speeds up the quantum evolution, whereas at comparatively higher temperatures, increasing acceleration slows down the quantum evolution. We also observe that this change in behavior begins at approximately $T \geq 1~\mathrm{K}$. Furthermore, for a high background temperature of about $300~\mathrm{K}$, as shown in Fig.~\ref{fig:thermalcase}(d), the corresponding acceleration scale is further reduced to approximately $10^{17}~\mathrm{m/s^2}$. Thus, increasing the temperature shifts the onset of the acceleration-induced effect to a lower acceleration regime. If we further increase the temperature, we can't see any significant changes. These results indicate that the background temperature has a significant influence on the QSLT and enhances the visibility of acceleration-related effects. As a result, the acceleration required to observe these effects is reduced compared to the zero-temperature case. Nevertheless, the required accelerations remain far beyond current experimental capabilities.

\begin{figure}[http!]
\centering
\includegraphics[width=0.49\columnwidth]{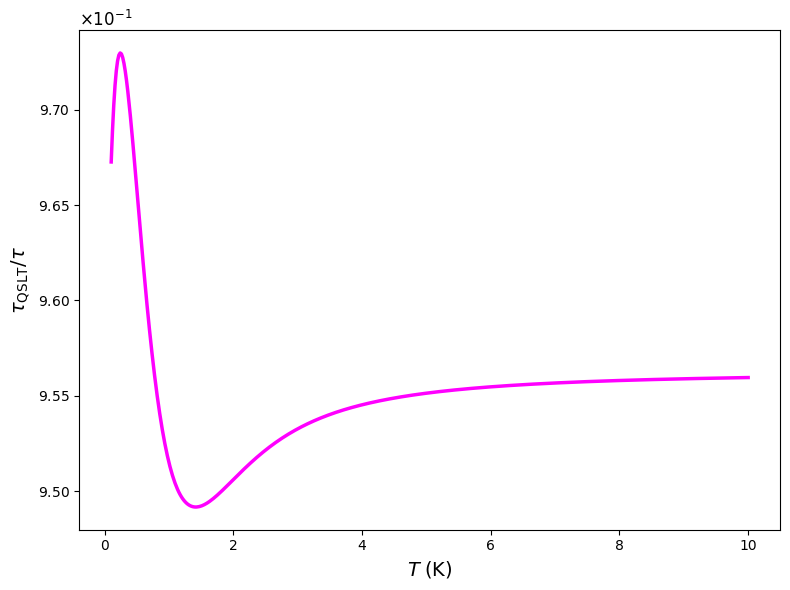}
\hfill
\includegraphics[width=0.49\columnwidth]{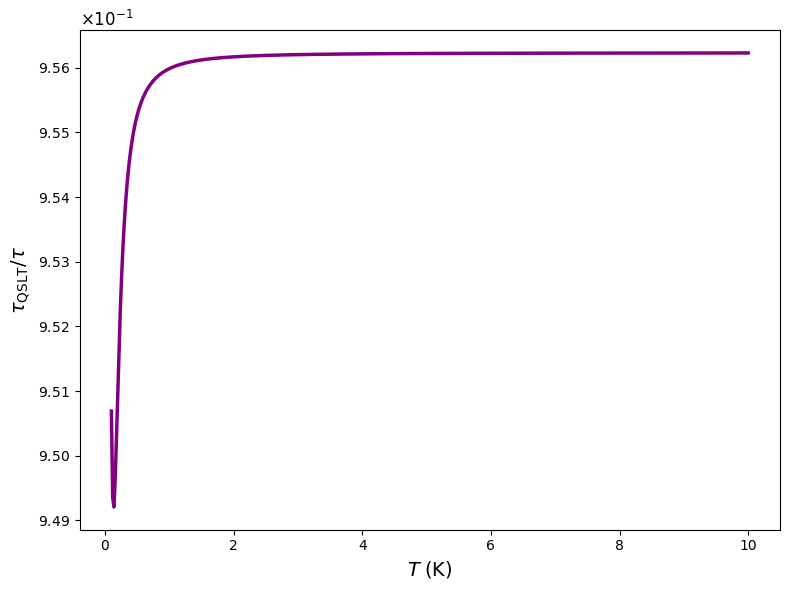}
\caption{Variation of  the normalized quantum speed limit time, $\tau_{\mathrm{QSLT}}/\tau$ as a function of temperature. The left panel corresponds to $a=10^{12}\,\mathrm{m/s^2}$, while the right panel corresponds to $a=10^{19}\,\mathrm{m/s^2}$.}
\label{fig:comparison}
\end{figure}

We mentioned earlier that the QSLT contains exchange symmetry $\beta_U\leftrightarrow \beta$. Therefore we expect that the behavior of the QSLT as a function of acceleration at fixed background temperature is qualitatively similar to its behavior as a function of temperature at fixed acceleration. If this is so, then the distinctive feature in QSLT with respect to acceleration can be interpreted as the indirect hallmark of Unruh effect. Looking at this aspect we now like to check this qualitative similarity by studying the variation of the QSLT with background temperature (see Fig.~\ref{fig:comparison}) at both low and high accelerations and  the comparing the results with the acceleration-dependent plots, given in Fig.\ref{fig:thermalcase}. If the same qualitative behavior is observed in both cases, it would support the interpretation that the changes in the QSLT arise from the Unruh effect.
From Fig. \ref{fig:comparison}, we observe that at low acceleration (corresponding to a low Unruh temperature), the QSLT decreases with increasing background temperature, which is similar to the behavior shown in Fig.~\ref{fig:thermalcase}(b). In contrast, at high acceleration (corresponding to a high Unruh temperature), the QSLT increases with increasing temperature, consistent with the behavior observed in Fig.~\ref{fig:thermalcase}(c). These results suggest that the acceleration, which corresponds to the Unruh temperature, influences the QSLT qualitatively in a similar manner to that of a real thermal background.
Therefore, the observation of the nontrivial variation of QSLT around a particular acceleration value, as shown in Fig. \ref{fig:thermalcase}, can be regarded as an indirect observation o Unruh effect. However we saw that such is achievable for the present scenario at a very large acceleration.

\section{Spacetime with reflecting boundaries} \label{Refl}
In previous attempts, it has been observed that the indirect detection of Unruh effect (like through Berry phase \cite{Barman:2024jpc}) can be done considerably less value of acceleration when the system is kept in presence of reflecting boundaries for the background field modes. In this situation the atom is moving parallel to the plane of the boundary. For instance if the atom is moving along $x$-axis, then the boundary is on the $x-y$ plane. Further entanglement harvesting between two similar detectors can be enhanced \cite{Barman:2023wkr}. Following this observation and considering the fact that QSLT can be a promising way to achieve the goal, in this section, we focus on the study of QSLT for an accelerated particle in the presence of single and double reflecting boundaries.

\subsection{Analytical expression}\label{Refl1}
The general expression for the quantum speed limit time is given by (\ref{eq10}). However, the coefficients \(A\) and \(B\) are modified because the two-point correlation function differs in the presence of boundaries. To obtain the expressions for \(A\) and \(B\) in the single-boundary case, we first calculate these coefficients for the double-boundary configuration. The corresponding expressions for a single boundary can then be obtained by considering the appropriate limit of the double-boundary result. The detector trajectories are given by (\ref{eq14}). Now restoring $c$ in the expressions we write them below in terms of the detector's proper time ($\tau$) as: 
\begin{equation}\label{19}
t=\frac{c}{a}\sinh\!\left(\frac{a\tau}{c}\right), \,\,\
x=\frac{c^{2}}{a}\cosh\!\left(\frac{a\tau}{c}\right), \,\,\
y=0, z=z_{0}~.
\end{equation}
Two mirrors are located at \(z=0\) and \(z=L\) with \(0<z_{0}<L\). 

\begin{widetext}
The Wightman function in the presence of two infinitely extended mirrors is given by \cite{Birrell:1982ix,Barman:2023wkr}
\begin{equation}\label{eq20}
    G_{W}(\Delta\tau)
=
-\frac{\hbar}{4\pi^{2}c}
\sum_{n=-\infty}^{\infty}
\left[
\frac{1}
{(c\Delta t-i\epsilon)^{2}-\Delta x^{2}-(2Ln)^{2}}
-
\frac{1}
{(c\Delta t-i\epsilon)^{2}-\Delta x^{2}-(2z_{0}-2Ln)^{2}}
\right].
\end{equation}
Using this one finds the coefficients $A, B$ as
\begin{equation}\label{eq21}
 A=\frac{\kappa a}{16\pi c}
\coth\!\left(\frac{\pi\omega_{0}c}{a}\right)
\sum_{n=-\infty}^{\infty}
\Big[
J(Ln)-J(Ln+z_{0})
\Big], \hspace{0.5cm}
B=\frac{A}{\coth\!\left(\dfrac{\pi\omega_{0}c}{a}\right)}.
\end{equation}
Here $J(u)$ and $\kappa$ are 
\begin{equation}
J(u)=\frac{\sin\!\left(\frac{2\omega_{0}c}{a}
\sinh^{-1}\!\left|\frac{au}{c^{2}}\right|\right)}
{\left|\dfrac{au}{c^{2}}\right|\left[\left(\dfrac{au}{c^{2}}\right)^{2}+1\right]^{\frac{1}{2}}}, \hspace{0.5cm}
\kappa=\frac{2 g^{2}}{\hbar c^{3}}.
 \end{equation}
 The explicit steps to find the above values are presented in Appendix \ref{App4} (see also \cite{Barman:2024jpc}). Remember that $n=0$ term of the above corresponds to the single boundary scenario.  
\end{widetext}

\subsection{Single reflecting boundary}\label{Refl2}
\begin{figure}[h!]
\centering
\includegraphics[width=\columnwidth]{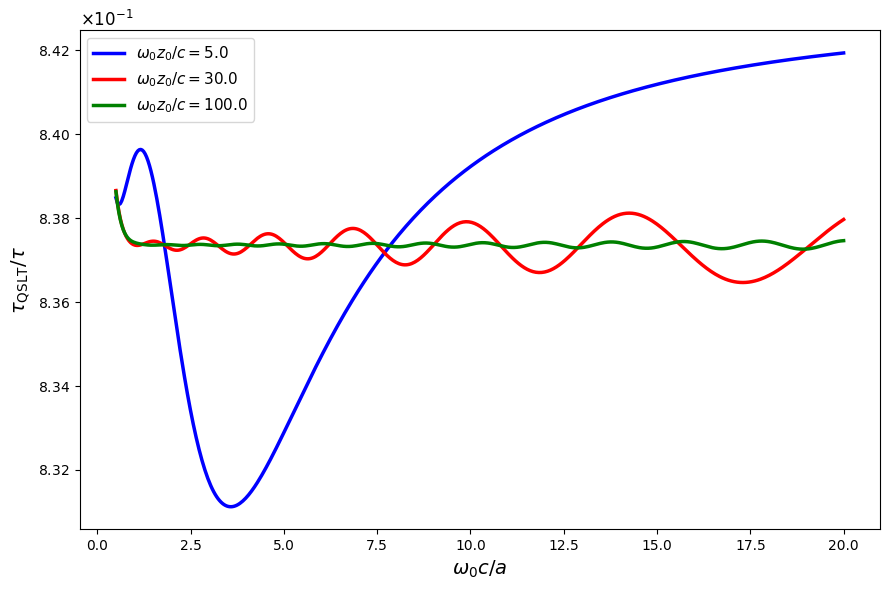}
\caption{Variation of the normalized quantum speed limit time, $\tau_{\mathrm{QSLT}}/\tau$, as a function of the acceleration $a$ ($\mathrm{m/s^2}$) in the presence of a single boundary, for different values of the dimensionless parameter $\omega_0 z_0/c = 5.0,\;30.0,$ and $100.0$. The remaining parameters are fixed at $\theta=\pi/4$, $\tau=10^{-9}\,\mathrm{s}$, $g=10^{-4}$, and $\omega_0 = 3 \times 10^9\,\mathrm{Hz}$.
}
\label{fig:SINGLE1}
\end{figure}

\begin{figure}[h!]
\centering
\includegraphics[width=\columnwidth]{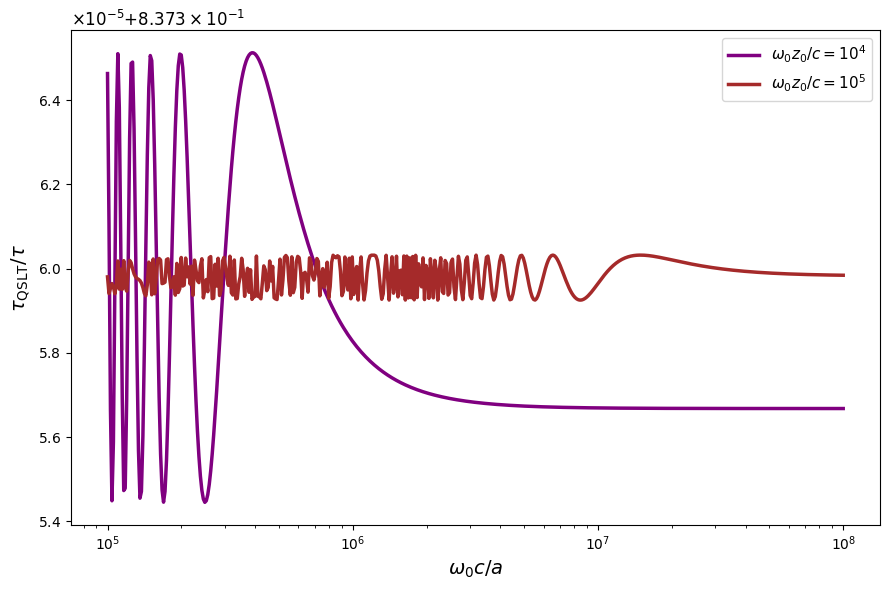}
\caption{Variation of the normalized quantum speed limit time, $\tau_{\mathrm{QSLT}}/\tau$, as a function of the acceleration $a$ ($\mathrm{m/s^2}$) in present of single boundary, for two different values of the dimensionless parameter $\omega_0 z_0/c = 10^4$ and $10^5$. The remaining parameters are fixed at $\theta=\pi/4$, $\tau=10^{-9}\,\mathrm{s}$, $g=10^{-4}$, and $\omega_0 = 3 \times 10^9 \mathrm{Hz}$.}
\label{fig:SINGLE2}
\end{figure}

For the single-mirror setup, we retain only the \(n=0\) term in the corresponding expressions. The coefficients $A$ and $B$ are then given by
\begin{equation}\label{eq23}
A=\frac{\kappa a}{16\pi c}\coth\!\left(\frac{\pi\omega_{0}c}{a}\right)\left[\frac{2\omega_{0}c}{a} -\frac{\sin\!\left(\frac{2\omega_{0}c}{a}
\sinh^{-1}\!\left|\frac{az_{0}}{c^{2}}\right|\right)}
{\left|\dfrac{az_{0}}{c^{2}}\right|\left[\left(\dfrac{az_{0}}{c^{2}}\right)^{2}+1\right]^{\frac{1}{2}}}
\right]~;
\end{equation}
and
\begin{equation}\label{eq24}
B=\frac{\kappa a}{16\pi c}
\left[\frac{2\omega_{0}c}{a}-\frac{\sin\!\left(
\frac{2\omega_{0}c}{a}
\sinh^{-1}\!\left|\frac{az_{0}}{c^{2}}\right|
\right)}
{\left|\dfrac{az_{0}}{c^{2}}\right|\left[\left(\dfrac{az_{0}}{c^{2}}\right)^{2}+1
\right]^{\frac{1}{2}}}
\right]~,
\end{equation}
respectively.
For the numerical analysis, we consider the atomic frequency as $\omega_0 \sim 3~\mathrm{GHz}$. Then following the previous discussion, we choose $g^2/(\hbar c^3) = 10^2$ for the evolution time $\tau = 10^{-9}~\mathrm{s}$. To study the behavior of the QSLT, we introduce the dimensionless parameters $(\omega_0 c/a)$ and $(\omega_0 z_0/c)$, representing the acceleration and distance parameters, respectively. We then plot $\tau_{\mathrm{QSLT}}/\tau$ as a function of $(\omega_0 c/a)$ in Fig. \ref{fig:SINGLE1} and Fig. \ref{fig:SINGLE2} for different values of $(\omega_0 z_0/c)$.

In Fig.~\ref{fig:SINGLE1},  we have taken low values of $(\omega_0 z_0/c)$, while Fig. \ref{fig:SINGLE2} corresponds to that of higher values. Here one observes a fluctuating behavior of $\tau_{QSLT}/\tau$ as a function of $(\omega_0 c/a)$ for different values of the parameter $(\omega_0 z_0/c)$. However, this behavior disappears beyond a certain value of $(\omega_0 c/a)$, after which the curves become nearly flat. Therefore the end or starting of the fluctuation in the QSLT due to the change of acceleration can be interpreted as the distinguishing effect of acceleration of the atom. We notice that increment of $(\omega_0 z_0/c)$ shifts the fluctuating region toward higher values of $(\omega_0 c/a)$, allowing the acceleration-induced effects to remain visible at lower accelerations. However the amplitude of the oscillation is low for high $(\omega_0 z_0/c)$.

In Fig.~\ref{fig:SINGLE2}, we consider two different values of the dimensionless distance parameter $(\omega_0 z_0/c)$, namely $10^4$ and $10^5$, which correspond to $z_0 = 1~\mathrm{km}$ and $z_0 = 10~\mathrm{km}$, respectively. For the case $z_0 = 10~\mathrm{km}$, we observe that the oscillatory behavior persists up to approximately $(\omega_0 c/a) \sim 10^7$, which corresponds to an acceleration of about $a \sim 10^{11}~\mathrm{m/s^2}$. This suggests that, if a particle can be accelerated uniformly at around $10^{11}~\mathrm{m/s^2}$, a noticeable change in the QSLT may be observed. So we see that presence of boundary indeed helping in reducing the required values of acceleration to observe the effect. The validity of the chosen parameter values within the range of acceleration taken here is studied in Appendix \ref{App2}.

\subsection{Double reflecting boundary}\label{Refl3}
Two reflecting boundaries are at $z=0$ and $z=L$ and the atom is moving along $x$-axis with a fixed value $z=z_0$ within the region between these boundaries. The coefficients $A$ and $B$ are then given by (\ref{eq21}). In these expressions, the summation over $n$ runs from $-\infty$ to $+\infty$. However, for large values of $n$, $a(Ln+z_0)$ becomes approximately equal to $aLn$, and therefore $J(Ln)-J(Ln+z_0)\approx 0$. Hence, the contribution from large values of $n$ is negligible.
For numerical calculations, we truncate the infinite summation by introducing lower and upper cutoffs and choose $n= - 10^{8}$ to $n= 10^{8}$. We observe this to be sufficient as further inclusion of terms effects a negligible change in the numerical values (a justification on this has been provided in Appendix \ref{App3}). We then use these coefficients to study the behavior of the QSLT as a function of $\omega_0 c/a$ for different values of $L\omega_0/c$ and different choices of $\omega_0 z_0/c$. Here also We observe an fluctuating behavior that persists in the low-acceleration region. We next focus on this low-acceleration region by choosing suitable values of $L\omega_0/c$ and $\omega_0 z_0/c$.

\begin{figure}[h!]
\centering
\includegraphics[width=\columnwidth]{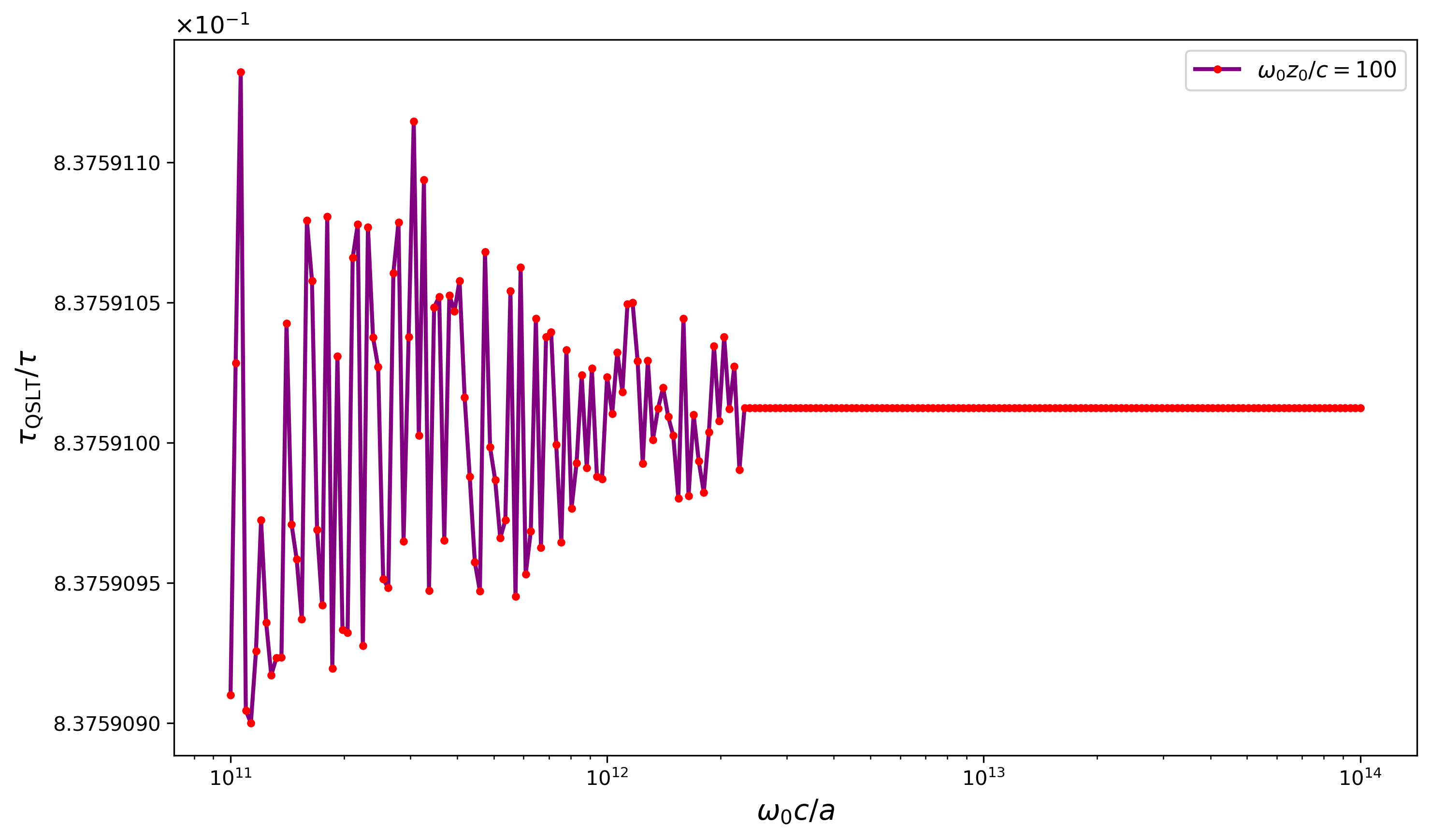}
\caption{Variation of the normalized quantum speed limit time, $\tau_{\mathrm{QSLT}}/\tau$, as a function of the acceleration parameter $\omega_0 c/a$ for a double-boundary configuration with $\omega_0 z_0/c=250$ and fixed $L\omega_0/c=1000$. The yellow dots represent the values obtained from Eq.~(\ref{eq10}), while the brown line connects these points. The remaining parameters are fixed at $\theta=\pi/4$, $\tau=10^{-9},\mathrm{s}$, $g=10^{-4}$, and $\omega_0=3\times10^9,\mathrm{Hz}$.
}
\label{fig:Double1}
\end{figure}
\begin{figure}[h!]
\centering
\includegraphics[width=\columnwidth]{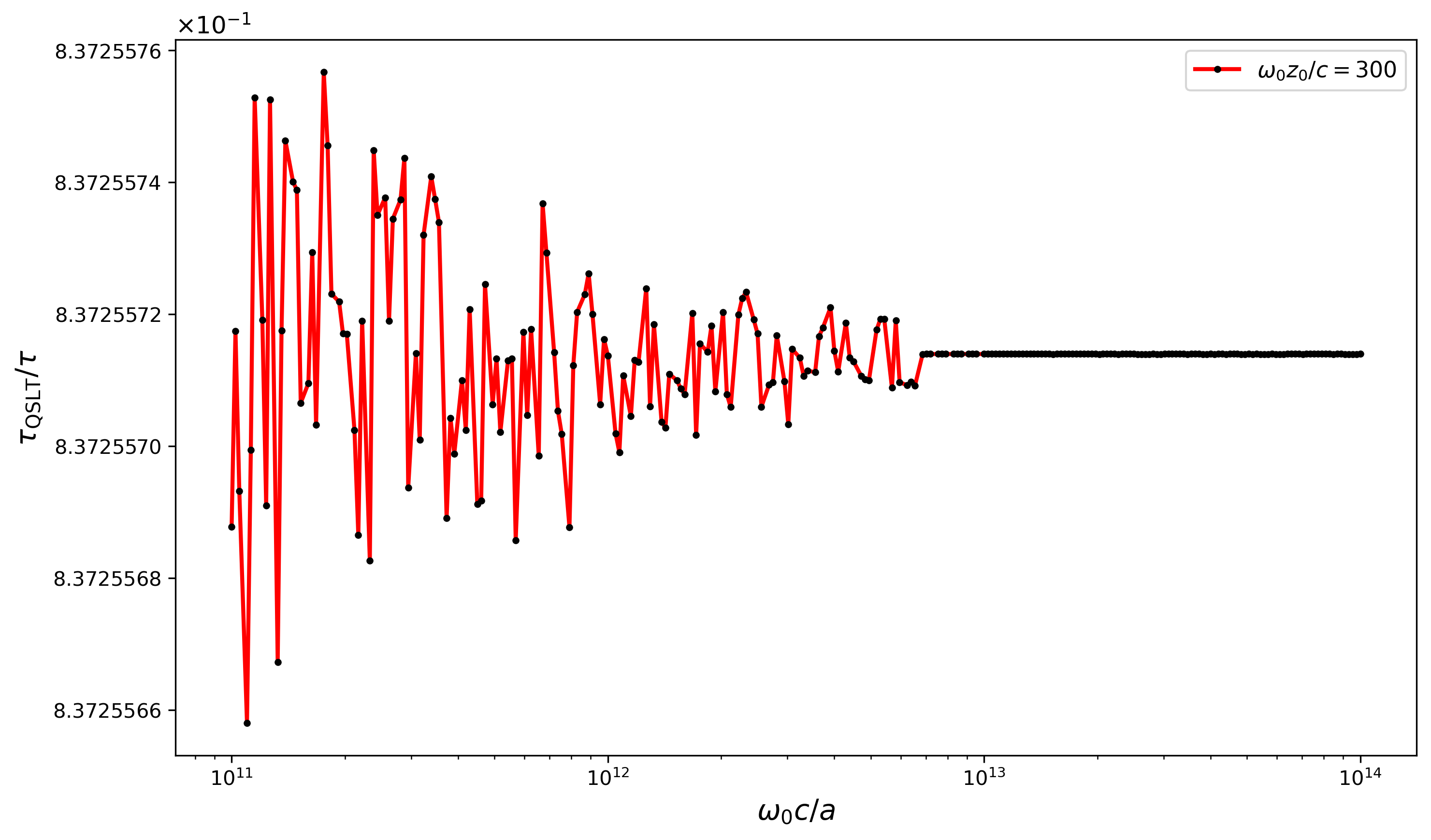}
\caption{Variation of the normalized quantum speed limit time, $\tau_{\mathrm{QSLT}}/\tau$, as a function of the acceleration parameter $\omega_0 c/a$ for a double-boundary configuration with $\omega_0 z_0/c=300$ and fixed $L\omega_0/c=3000$. The black dots represent the values obtained from Eq.~(\ref{eq10}), while the red line connects these points. The remaining parameters are fixed at $\theta=\pi/4$, $\tau=10^{-9},\mathrm{s}$, $g=10^{-4}$, and $\omega_0=3\times10^9,\mathrm{Hz}$.
}
\label{fig:Double2}
\end{figure}
In Fig.~\ref{fig:Double1} and Fig.~\ref{fig:Double2}, we observe that the fluctuating behavior of the QSLT can persist until larger values of the dimensionless acceleration parameter $(\omega_0 c/a)$, which is increased by increasing the mirror separation. For $L\omega_0/c = 1000$ (i.e. $L=100~\mathrm{m}$) and $\omega_0 z_0/c = 250$ (i.e. $z_0=25~\mathrm{m}$), the fluctuations persist up to $(\omega_0 c/a)\sim10^{12}$, corresponding to an acceleration of approximately $10^{6}~\mathrm{m/s^2}$. Increasing the parameters further to $L\omega_0/c = 3000$ (i.e., $L=300~\mathrm{m}$) and $\omega_0 z_0/c = 300$ (i.e., $z_0=30~\mathrm{m}$), the fluctuating behavior extends up to $(\omega_0 c/a)\sim10^{13}$, corresponding to an acceleration of about $10^{5}~\mathrm{m/s^2}$. Although the amplitude of the oscillations is very small, the corresponding change in the QSLT remains within the experimentally accessible time scale (in particular, time resolutions down to the attosecond regime have been experimentally demonstrated~\cite{Hentschel2001}). The justification for the above chosen parameter values is  provided in Appendix~\ref{App2}. 

This suggests that if a qubit system is accelerated with these fixed parameter arrangements and the acceleration is increasing from small values, then a noticeable change in the behavior of the QSLT going to start observing with the minimum acceleration $a \sim 10^{5}~\mathrm{m/s^2}$. 
We can extend the fluctuating behavior by increasing $L\omega_0/c$. However, there are two main limitations. First, very large values of $L\omega_0/c$ may not be experimentally suitable. Second, in the high $(\omega_0 c/a)$ region, the amplitude of the fluctuations decreases, making the corresponding changes in $\tau_{\mathrm{QSLT}}/\tau$ difficult to detect.

\section{Proposal to measure QSLT}\label{Exp}
QSLT cannot be measured directly because it is not the time measured by an external clock. Instead, it is an intrinsic property of the quantum evolution of the system. Therefore, we propose an indirect method based on {\it quantum state tomography} (QST) (for review on QST, see \cite{Altepeter2004}). In this method, a large number of identical qubits are prepared in the same initial state and allowed to evolve under uniform acceleration for a time $\tau$. The density matrix at time $\tau$ is then reconstructed using QST. From the reconstructed density matrix and the known initial state, the QSLT is calculated using its theoretical expression. Comparing this value with that of the theoretical prediction from our evolved state theoretically obtained provides a way to test our results.

To understand this method in more detail, we first briefly discuss QST. The main goal of QST is to reconstruct the density matrix of an unknown quantum state from a series of measurements performed on an ensemble of identically prepared particles. In practice, the exact density matrix cannot be reconstructed because an infinite number of measurements would be required to completely eliminate statistical errors. In the present work, however, we neglect these statistical errors and assume ideal (exact) quantum state tomography.

Any single-qubit density matrix $\hat{\rho}$ can be uniquely represented by three Stokes parameters $(S_1,S_2,S_3)$ as
\begin{equation}\label{eq25}
\hat{\rho}=\frac{1}{2}\sum_{i=0}^{3}S_i\sigma_i,
\end{equation}
where $\sigma_i$ are the Pauli matrices. The Stokes parameters are given by
\begin{equation}
S_i=\mathrm{Tr}(\hat{\rho}\,\sigma_i)~,
\end{equation}
with $\sigma_0 = 1$ and hence $S_0=1$.
Therefore, if the Stokes parameters $S_i$ can be measured experimentally, the density matrix can be reconstructed directly using Eq.~(\ref{eq25}).
The Stokes parameters are measured with respect to the states $|\phi_1\rangle=\frac{|0\rangle+|1\rangle}{\sqrt{2}}$, $|\phi_2\rangle=\frac{|0\rangle+i|1\rangle}{\sqrt{2}}$, and $|\phi_3\rangle=|0\rangle$, together with their corresponding orthogonal complements $|\phi_1^\perp\rangle$, $|\phi_2^\perp\rangle$, and $|\phi_3^\perp\rangle$. Physically, each of these parameters corresponds to the outcome of a specific pair of projective measurements \cite{Altepeter2004}:
\begin{align}
S_0 &= P_{|\phi_3\rangle}+P_{|\phi_3^{\perp}\rangle}, \\
S_1 &= P_{|\phi_1\rangle}-P_{|\phi_1^{\perp}\rangle}, \\
S_2 &= P_{|\phi_2\rangle}-P_{|\phi_2^{\perp}\rangle}, \\
S_3 &= P_{|\phi_3\rangle}-P_{|\phi_3^{\perp}\rangle},
\end{align}
where $P_{|\phi_{\mu}\rangle}$ is the probability of measuring the state $|\phi_{\mu}\rangle$ (here $\mu = 1,2,3$). Since
\begin{equation}
P_{|\phi_{\mu}\rangle}+P_{|\phi^{\perp}_{\mu}\rangle}=1,
\end{equation}
the above expressions can be simplified for a single qubit as
\begin{equation}
P_{|\phi_{\mu}\rangle}-P_{|\phi^{\perp}_{\mu}\rangle}
=
2P_{|\phi_{\mu}\rangle}-1,
\end{equation}
where $|\phi^{\perp}_{\mu}\rangle$ denotes the state orthogonal to $|\phi_{\mu}\rangle$.
The protocol for reconstructing the density matrix is as follows: (i) prepare a single qubit in the initial state $|\phi_0\rangle$, (ii) allow the qubit to evolve for a time $\tau$, and (iii) perform projective measurements to determine the Stokes parameters $(S_1,S_2,S_3)$. Using these measured Stokes parameters, the density matrix is reconstructed through Eq.~(\ref{eq25}).

In our case, the quantum state tomography protocol begins by preparing a large ensemble of identical qubits in the same initial state $\rho(0)=|\psi(0)\rangle\langle\psi(0)|$. Each qubit is then subjected to the same uniform acceleration and allowed to evolve for a proper time $\tau$. Our main objective is to experimentally reconstruct the evolved density matrix $\rho_{\mathrm{ex}}(\tau)$. 
For this, we perform projective measurements on the evolved qubits, from which the Stokes parameters can be determined experimentally. Since a single projective measurement cannot determine the complete quantum state, the ensemble is divided into three identical groups. The first group is measured in the computational basis $\{|0\rangle,|1\rangle\}$ to determine $\langle\sigma_z\rangle$($S_3$). The second group is subjected to a $\pi/2$ pulse before measurement, which effectively changes the measurement basis to $\{(|0\rangle+|1\rangle)/\sqrt{2},(|0\rangle-|1\rangle)/\sqrt{2}\}$ and allows us to determine $\langle\sigma_x\rangle$($S_1$). Similarly, a phase-shifted $\pi/2$ pulse is applied to the third group, giving the measurement basis $\{(|0\rangle+i|1\rangle)/\sqrt{2},(|0\rangle-i|1\rangle)/\sqrt{2}\}$ and allowing us to determine $\langle\sigma_y\rangle$($S_2$). Each measurement is repeated many times to estimate the corresponding probabilities, and in the present analysis we neglect the associated statistical errors. Finally, the three expectation values $\langle\sigma_x\rangle$, $\langle\sigma_y\rangle$, and $\langle\sigma_z\rangle$ determine the Stokes parameters of the evolved state, from which the density matrix is reconstructed as
\begin{equation}
\rho_{\mathrm{ex}}(\tau)
=
\frac{1}{2}
\left[
I
+\langle\sigma_x\rangle\sigma_x
+\langle\sigma_y\rangle\sigma_y
+\langle\sigma_z\rangle\sigma_z
\right].
\end{equation}

The tomography procedure is repeated for different evolution times between $0$ and $\tau$, which allows us to reconstruct $\rho_{\mathrm{ex}}(\tau)$ at each intermediate time and, consequently, to evaluate the time-averaged speed term appearing in the denominator of the QSLT expression. The experimentally reconstructed state $\rho_{\mathrm{ex}}(\tau)$ and the corresponding time-averaged speed are then substituted into the theoretical expression for the QSLT [Eq.~(\ref{eq8})] to obtain the experimental value $\tau_{\mathrm{QSLT}}^{\mathrm{ex}}$. This procedure is repeated for different values of the acceleration, and the resulting $\tau_{\mathrm{QSLT}}^{\mathrm{ex}}$ is plotted as a function of acceleration. If the experimentally obtained curve shows the same qualitative behavior as the theoretical prediction given by Eq.~(\ref{eq10}), this would provide experimental support for the acceleration-induced signature associated with the Unruh effect.


\section{Conclusions and outlook}\label{Concl}
Apart from non-relativistic analysis, the relativistic effects on the information processing appear to be very interesting and active area for the last two decades. In this direction, Unruh effect plays crucial role. QSLT has been an important concept in quantum information theory. Here we studied the properties of QSLT of a uniformly accelerated Unruh-DeWitt detector which is interacting with background real massless scalar field. Within the open quantum system formalism, under the application of Markov process, we investigated the role of acceleration on the speed of the detector's quantum processes. We considered three scenarios -- (i) the field is in Minkowski spacetime without any reflecting boundary but at a finite temperature, (ii) the presence of single boundary with the detector is accelerating parallel to the plane of it, and (iii) the presence of two boundaries with the  detector is moving between them along the parallel direction. This investigation not only explores the relativistic effects on the QSLT, but also revels the effects of the boundary for the environment. As far as we are aware, this is the first attempt done on the Unruh-DeWitt detector. 

In all the cases, depending upon the parameters and the ranges of acceleration, QSLT is either increasing or decreasing due to the acceleration of the detector. This means the quantum transition from the initial state to the final state in an accelerated detector can be slower or faster within certain ranges of acceleration. Crucially, the first scenario, i.e. case (i), revels the equivalent roles played by temperature of the field and the acceleration of the detector. Such an observation indicates that the Unruh thermalization must gives its signature in QSLT. Therefore any non-trivial feature of QSLT as a function of acceleration can be an indirect footprint of Unruh effect. Based on this idea, we further investigated the effects of boundaries which act as the reflecting barrier for the background field modes. One distinguishable feature appeared. Both for single boundary and double boundaries, QSLT starts fluctuating within a measurable range after a critical value of the acceleration. Fluctuation is more vigorous towards larger values of acceleration. We proposed that the appearance of this fluctuation can be used for an indirect detection of Unruh effect. Estimation of parameters showed that the critical values of the acceleration is much less for the double boundary situation compared to the single one. For the observable range of fluctuation the required critical acceleration, depending upon the reasonable values of the other parameters,  can be as small as $a\sim 10^5 {\text{m/s$^2$}}$. Note that in all analysis we set the evolution time $\tau$ as the minimum possible value $\tau = 10^{-9} \text{s}$. However, one can choose larger value. In this case the characteristic nature of  $\tau_{\text{QSLT}}$ as a function of acceleration does not change (as an example see Appendix \ref{App5}, where the evolution time is considered as $\tau = 1 {\text{s}}$). 

In this connection, we like to mention that the previous proposed models to visualize Unruh effect suggested the required acceleration as $a\sim 10^8$ -- $10^9 \text{m/s}^2$ (e.g. see \cite{Lochan:2019osm}). Further, the Pancharatnam-Berry phase estimation proposal within our two-boundary scenario yielded again a similar range \cite{Barman:2024jpc}. To observe the minimum measurable phase one requires $L= 3 ~\text{km}$, $z_0 = 33 ~\text{m}$ and $a\sim 10^8 \text{m/s}^2$. However, the present analysis through QSLT showed better prospective. For the observable range of QSLT one needs $L= 300~\text{m}$, $z_0 = 30~\text{m}$ and $a\sim 10^5\text{m/s}^2$. Therefore, for the indirect observation of Unruh effect, QSLT seems to be more effective than the phase estimation. We also discuss a possible sketch to measure the QSLT.

Finally, it should be noted that the present model is far from the reality. In particular, inclusion of the monopole detector and real scalar field in the investigation is not feasible for practical purposes. However, most studies in the context of relativistic quantum information processes are done considering this simplified model. This is because such an approximation helps to provide an analytical analysis, thereby illuminating the underlying physics and hence increases our understandings. Furthermore, the monopole detector can be a good approximated version of dipole atomic detector. Our present study is inspired by these facts. Looking at the promising future of the QSLT to detect Unruh effect, we like to build a better model which can be implemented in the laboratory. Particularly, a model, built out of an atomic dipole interacting with background electric field, will be more fruitful. This we will address in our forthcoming work \cite{Fwork}. On the other hand the Einstein's equivalence principle promotes these observations likely to happen for a static detector stationed near the black hole horizon. However, this classical equivalence principle may not be consistent for a quantum system. Therefore a detailed investigation in presence of black hole will be interesting.



\begin{widetext}
\appendix
\section*{\Huge{Appendices}}
\section{Derivation of Eq. (\ref{eq10})} \label{App1}
The explicit expression for $\tau_{\mathrm{QSLT}}/\tau$ can be obtained by substituting Eq.~(\ref{eq7}) into Eq.~(\ref{eq8}). To evaluate Eq.~(\ref{eq8}), we calculate it in two parts: the numerator and the denominator. Let's start with the numerator part. The numerator of Eq.~(\ref{eq8}) is the Hilbert--Schmidt distance between the initial state $\rho_c(0)$ and the evolved state $\rho_c(\tau)$. Here this is being denoted as
$\mathcal{D} = \left\| \rho_c(0) - \rho_c(\tau) \right\|_{\mathrm{HS}}$.
To proceed, we symbolize few quantities as 
$X = e^{-4A\tau}, \quad 
Y = e^{-2A\tau - i\Omega\tau}, \quad 
Y^* = e^{-2A\tau + i\Omega\tau}$.
Then in these notations one finds
\begin{equation}
\rho_c(0) - \rho_c(\tau)
=
\begin{bmatrix}
\cos^2\left(\frac{\theta}{2}\right)(1 - X)
+ \frac{A-B}{2A}(X - 1)
&
\frac{1}{2}\sin\theta (1 - Y)
\\[8pt]
\frac{1}{2}\sin\theta (1 - Y^*)
&
\cos^2\left(\frac{\theta}{2}\right)(X - 1)
- \frac{A-B}{2A}(X - 1)
\end{bmatrix}
\equiv \begin{bmatrix}
m_{11} & m_{12} \\
m_{21} & m_{22}
\end{bmatrix}~.
\label{A2}
\end{equation}
For the above matrix square of the Hilbert--Schmidt norm is given by
\begin{equation}
\|M\|_{\mathrm{HS}}^2
= |m_{11}|^2 + |m_{22}|^2 + |m_{12}|^2 + |m_{21}|^2.
\label{A3}
\end{equation}
Substitution of the explicit form of the matrix elements from (\ref{A2}) yields 
\begin{equation}
\begin{aligned}
\|M\|_{\mathrm{HS}}^2
&=
2 \left[
\cos^2\left(\frac{\theta}{2}\right)(1 - e^{-4A\tau})
+ \frac{A-B}{2A}(e^{-4A\tau}-1)
\right]^2
+\frac{1}{2}\sin^2\theta
\left(1 - 2e^{-2A\tau}\cos(\Omega\tau) + e^{-4A\tau}\right).
\end{aligned}
\end{equation}
Hence the Hilbert--Schmidt distance between the two states comes out to be
\begin{equation}
\mathcal{D} = \left\| \rho_c(0) - \rho_c(\tau) \right\|_{\mathrm{HS}}
= \sqrt{\|M\|_{\mathrm{HS}}^2}=\sqrt{
2\left(e^{-4A\tau}-1\right)^2 F^2
+\frac{1}{2}\left(1+e^{-4A\tau}-2e^{-2A\tau}\cos(\Omega\tau)\right)\sin^2\theta}~,
\label{A6}
\end{equation}
where $F$ is given by Eq. (\ref{11}). 

Next, we evaluate the denominator of Eq.~(\ref{eq8}). This quantity corresponds to the time-averaged speed of the quantum evolution from the initial state to the final state.
The time derivative of the density matrix is
\begin{equation}
\dot{\rho}_c(\tau') =
\begin{bmatrix}
\dot{m}_{11} & \dot{m}_{12} \\
\dot{m}_{21} & \dot{m}_{22}
\end{bmatrix}
\end{equation}
where
\begin{equation}
\begin{aligned}
\dot{m}_{11}
&= -4A e^{-4A\tau'}
\left[
\cos^2\left(\frac{\theta}{2}\right)
- \frac{A-B}{2A}
\right], \qquad
\dot{m}_{22}
= -\dot{m}_{11},
\\[6pt]
\dot{m}_{12}
&= \frac{1}{2}(-2A-i\Omega)\,
e^{-2A\tau'-i\Omega\tau'}\sin\theta,
\qquad
\dot{m}_{21}
=
\frac{1}{2}(-2A+i\Omega)\,
e^{-2A\tau'+i\Omega\tau'}\sin\theta .
\end{aligned}
\label{A7}
\end{equation}
The square of the Hilbert--Schmidt norm of this matrix is then given by
\begin{equation}
\begin{aligned}
\|\dot{\rho}_c(\tau')\|_{\mathrm{HS}}^2
&=
32A^2 e^{-8A\tau'}
\left[
\cos^2\left(\frac{\theta}{2}\right)
-\frac{A-B}{2A}
\right]^2+
\frac{1}{2}(4A^2+\Omega^2)e^{-4A\tau'}\sin^2\theta~.
\end{aligned}
\label{A10}
\end{equation}
Using this in Eq. (\ref{9}) we find the time-averaged Hilbert--Schmidt norm as
\begin{align}
\overline{\|\dot{\rho}_c(\tau')\|}_{\mathrm{HS}}
&=
\frac{1}{\tau}
\int_0^\tau
\Bigg[
32A^2 e^{-8A\tau'}
\left[
\cos^2\left(\frac{\theta}{2}\right)
-\frac{A-B}{2A}
\right]^2 +
\frac{1}{2}(4A^2+\Omega^2)e^{-4A\tau'}\sin^2\theta
\Bigg]^{1/2}
\, d\tau'.
\end{align}
Next taking the help of the standard integration result
\begin{align}
I &= \int_0^\tau \sqrt{C_1 e^{-8A\tau'} + C_2 e^{-4A\tau'}} \, d\tau'
\\
&=
\frac{1}{2A}
\Bigg\{
\frac{1}{2}\sqrt{C_1 + C_2}
+
\frac{C_2}{2\sqrt{C_1}}
\ln\!\left(
\sqrt{C_1} + \sqrt{C_1 + C_2}
\right)
\nonumber -
\frac{e^{-2A\tau}}{2}
\sqrt{C_1 e^{-4A\tau} + C_2}
\nonumber -
\frac{C_2}{2\sqrt{C_1}}
\ln\!\left(
\sqrt{C_1} e^{-2A\tau}
+
\sqrt{C_1 e^{-4A\tau} + C_2}
\right)
\Bigg\}~,
\end{align}
where the $C_1$ and $C_2$ coefficients are identified as 
\begin{equation}
C_1 = 32 A^2 \left(\cos^2\left(\frac{\theta}{2}\right) + \frac{B - A}{2A}\right)^2,\hspace{1cm}
C_2 = \frac{1}{2}  \; (4A^2 + \Omega^2) \, \sin^2\theta~,
\end{equation}
one can find the explicit expression of Eq. (\ref{9}). Finally using this and Eq. (\ref{A6}) we obtain Eq.~(\ref{eq10}).

\section{Validity of the chosen fixed parameters} \label{App2}
In Sec.~\ref{Free}, we chose a particular set of parameters to plot $\tau_{\mathrm{QSLT}}/\tau$. Similar values are being considered in Sec. \ref{Refl} as well. These chosen parameters must be consistent with our perturbation approximation employed in the calculation. One way to justify by investigating the first diagonal element $\rho_{11}$ of Eq.~(\ref{eq7}). Since $\rho_{11}$ represents a probability, it must remain non-negative. Also the other diagonal element is $\rho_{22}=1-\rho_{11}$, which must be positive and hence $\rho_{11}$ must be less than or equal to unity. Therefore, for the validity of the perturbative approximation, the chosen parameters must be such that $\rho_{11}$ satisfies $0 \leq \rho_{11} \leq 1$ throughout the entire acceleration range considered in our analysis.

We now plot $\rho_{11}$ as a function of acceleration, separately for the chosen parameters and range of acceleration taken in Sec. \ref{Free} and  Sec. \ref{Refl}. Fig. \ref{fig:probthermal} corresponds to Sec. \ref{Free} while Fig. \ref{fig:probsingle} and Fig. \ref{fig:probdouble} refer to single and double boundaries cases of Sec. \ref{Refl}, respectively.
\begin{figure}[!h]
\centering
\includegraphics[width=0.5\textwidth]{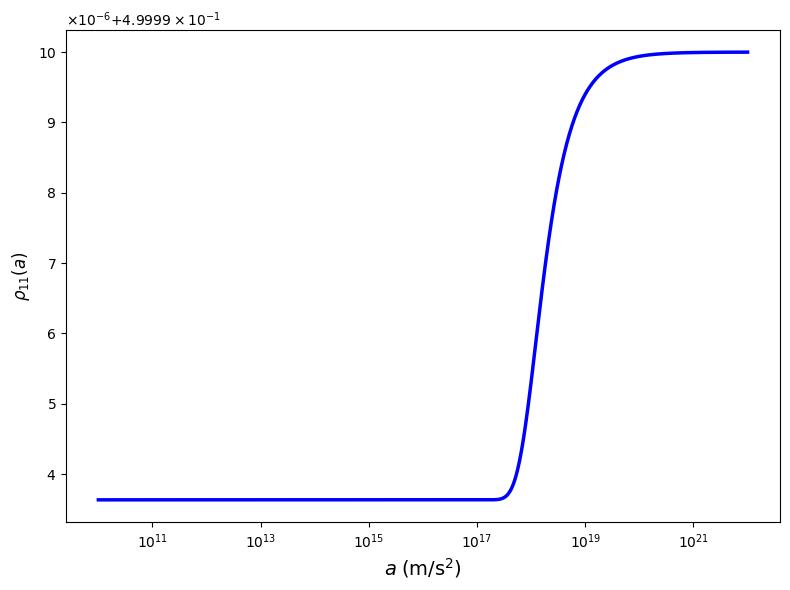}
\caption{Variation of the $\rho_{11}$ as a function of the acceleration. The remaining parameters are fixed at $\theta=\pi/4$, $\tau=10^{-9}\,\mathrm{s}$, $g=10^{-5}$, and $\omega_0 = 1 \times 10^9\,\mathrm{Hz}$.}
\label{fig:probthermal}
\end{figure}
\begin{figure}[!h]
\centering
\includegraphics[width=0.5\textwidth]{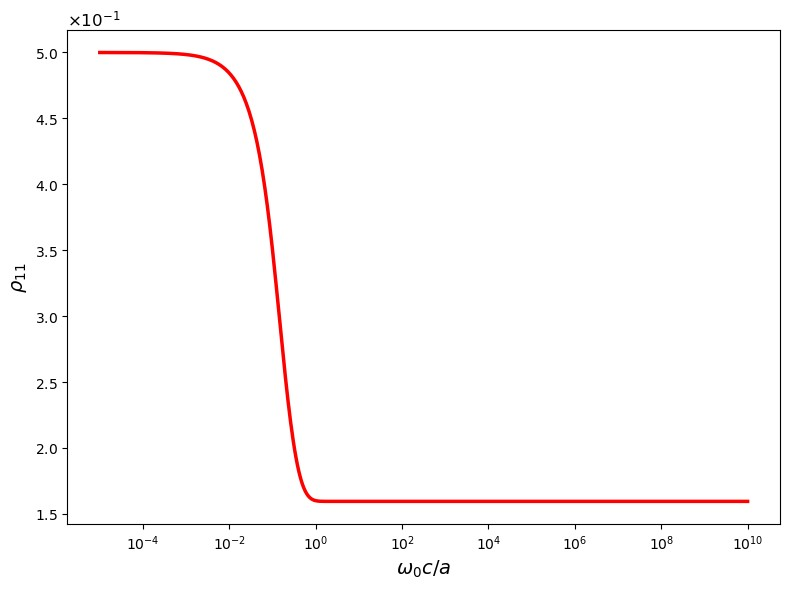}
\caption{Variation of the transition probability $\rho_{11}$ as a function of the dimensionless acceleration parameter $(\omega_0 c/a)$ for single boundary case. The remaining parameters are fixed at $\omega_0 z_0/c = 10^4$, $\theta=\pi/4$, $\tau=10^{-9}\,\mathrm{s}$, $g=10^{-4}$, and $\omega_0 = 3\times10^9\,\mathrm{Hz}$. Plot is shown entire acceleration range ($10^8$--$10^{23}\,\mathrm{m/s^2}$).}
\label{fig:probsingle}
\end{figure}
\begin{figure}[!h]
\centering
\includegraphics[width=0.5\textwidth]{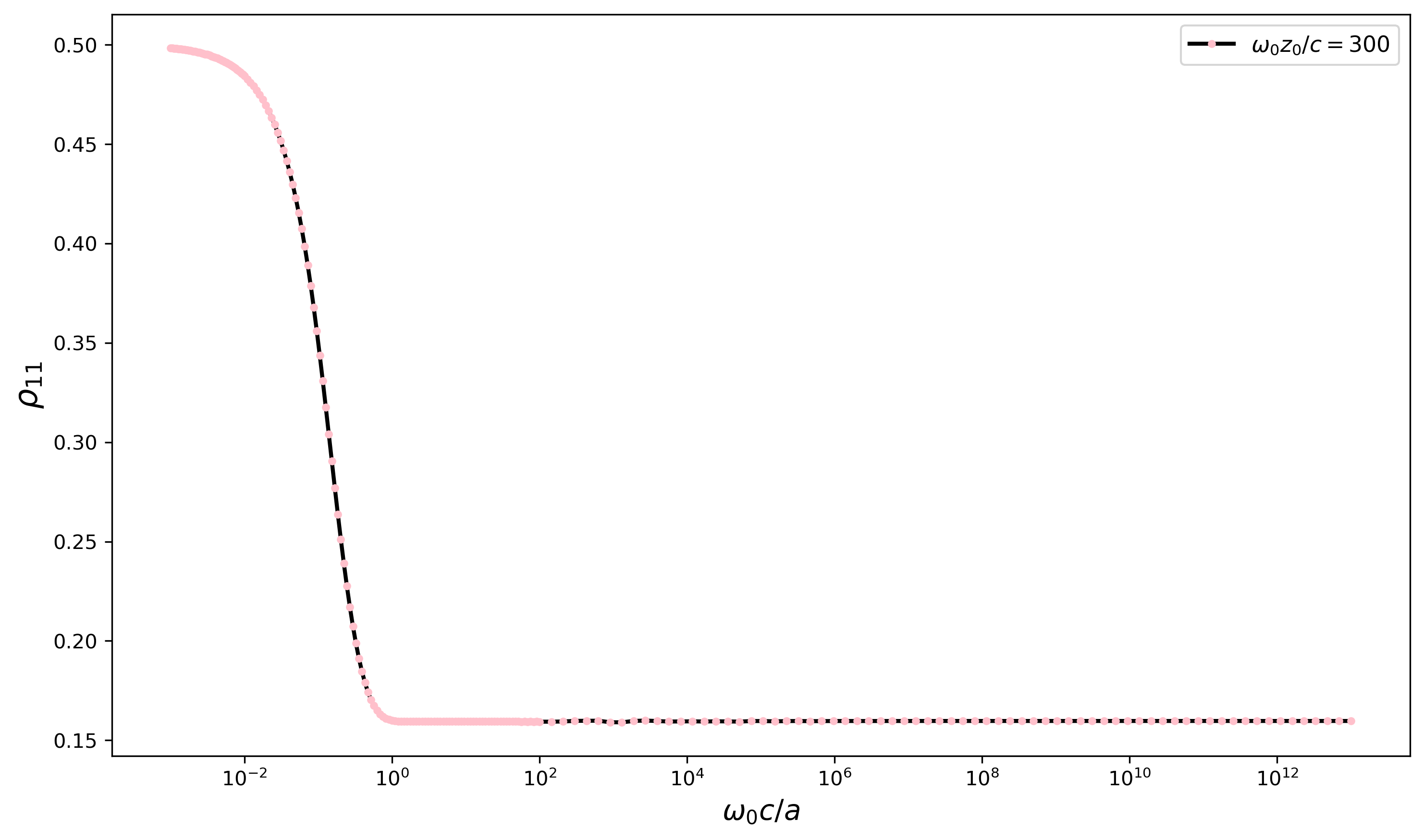}
\caption{Variation of the transition probability $\rho_{11}$ as a function of the dimensionless acceleration parameter $(\omega_0 c/a)$ for double boundary case. The remaining parameters are fixed at $\omega_0 z_0/c = 300$, $L\omega_0/c = 3000$, $\theta=\pi/4$, $\tau=10^{-9}\,\mathrm{s}$, $g=10^{-4}$, and $\omega_0 = 3\times10^9\,\mathrm{Hz}$.The pink dots represent the values obtained from Eq.~(\ref{eq10}), while the black line connects these points. Plot is shown entire acceleration range ($10^5$--$10^{21}\,\mathrm{m/s^2}$).}
\label{fig:probdouble}
\end{figure}
In each case, we observe that $\rho_{11}$ always remains within the range $0 \leq \rho_{11} \leq 1$. Therefore, the chosen set of parameters is physically consistent and supports the validity of the perturbative approach used in our analysis.

\section{Derivation of Eq. (\ref{eq21})}\label{App4}
Here, we present the explicit calculation of the coefficient $A$. To obtain it, we first transform the Wightman function given in Eq.~(\ref{eq20}) into Rindler coordinates. By substituting the coordinate transformation given in Eq.~(\ref{19}) into Eq.~(\ref{eq20}), we obtain
\begin{equation}\label{C1}
\begin{aligned}
G_{W}(\Delta\tau)
=-\frac{\hbar}{16\pi^{2}c}\sum_{n=-\infty}^{\infty}
\Bigg[\frac{1}{\dfrac{c^{4}}{a^{2}}\left(\sinh^{2}\!\left(\dfrac{a\Delta\tau}{2c}-i\epsilon\right)-\dfrac{a^{2}L^{2}n^{2}}{c^{4}}\right)}-\frac{1}{\dfrac{c^{4}}{a^{2}}
\left(\sinh^{2}\!\left(\dfrac{a\Delta\tau}{2c}-i\epsilon\right)-\dfrac{a^{2}(z_{0}+Ln)^{2}}{c^{4}}\right)}
\Bigg].
\end{aligned}
\end{equation}
Substituting the above in (\ref{eq5}) one can find $\gamma(\pm\omega_0)$. Since 
$G_{W}(\Delta\tau)$ is composed of two terms, $\gamma(\pm\omega_0)$ will have two terms to evaluate. Note that evaluation of one term will be enough to find the same for the other one. 
To proceed we denote 
\begin{equation}
K_{n1}^2 = \dfrac{a^{2}L^{2}n^{2}}{c^{4}}~; \,\,\,\  K_{n2}^2 = \dfrac{a^{2}(z_{0}+Ln)^{2}}{c^{4}}~.
\end{equation}
The first term of (\ref{C1}) then yields
\begin{equation}
\gamma_1(\omega_0)
=
-\frac{\hbar}{16\pi^{2}c}
\sum_{n=-\infty}^{\infty}
\int_{-\infty}^{\infty}
d\Delta \tau\,
\frac{e^{i\omega_0 \Delta \tau}}
{\dfrac{c^{4}}{a^{2}}\left(\sinh^{2}\!\left(\dfrac{a\Delta\tau}{2c}-i\epsilon\right)-K_{n1}^{2}\right)}~.
\end{equation}
Next, we introduce the dimensionless variables $S=\dfrac{a\Delta\tau}{2c}$ and $\alpha=\dfrac{2c\omega_0}{a}$. Using these substitutions, the above expression becomes
\begin{equation}\label{C3}
\gamma_1(\omega_0)
=
-\frac{\hbar a}{8\pi^{2}c^{4}}
\sum_{n=-\infty}^{\infty}
\int_{-\infty}^{\infty}
dS\,
\frac{e^{i\alpha S}}
{\sinh^{2}(S-i\epsilon)-K_{n1}^{2}}.
\end{equation}
We now focus on the integration part of Eq.~(\ref{C3}), which can be evaluated using the method of contour integration. The integral is given by
\begin{equation}
I=\int_{-\infty}^{\infty}
dS\,
\frac{e^{i\alpha S}}
{\sinh^{2}(S-i\epsilon)-K_{n1}^{2}}.
\end{equation}
We evaluate the above integral by extending it to the complex $S$-plane and considering the closed contour integral
\begin{equation}
I_C=
\oint_C
dS\,
\frac{e^{i\alpha S}}
{\sinh^{2}(S-i\epsilon)-K_{n1}^{2}},
\label{Ic}
\end{equation}
where the contour $C$ is chosen to enclose the poles of the integrand in the upper half-plane when $\alpha > 0$ and lower half plane when $\alpha < 0$.
Let $u_{n1}=\sinh^{-1}(\left|K_{n1}\right|)$. Then the poles of the integrand are
\begin{align}
S_1 &= u_{n1}+i\pi m+i\epsilon,\\
S_2 &= -u_{n1}+i\pi m+i\epsilon,
\end{align}
where $m=0,\pm1,\pm2,\cdots$. Now, let us consider the case $\alpha>0$, which corresponds to $+\omega_0$. In this case, we choose the contour in the upper half of the complex $S$-plane, as shown in Fig.~\ref{fig:upper}. From the figure, it is clear that only the poles lying in the upper half-plane are enclosed by the contour, while the poles in the lower half-plane, corresponding to $m=-1,-2,-3,\ldots$, are excluded.
\begin{figure}[!h]
\centering
\subfloat[]{
\includegraphics[width=0.45\textwidth]{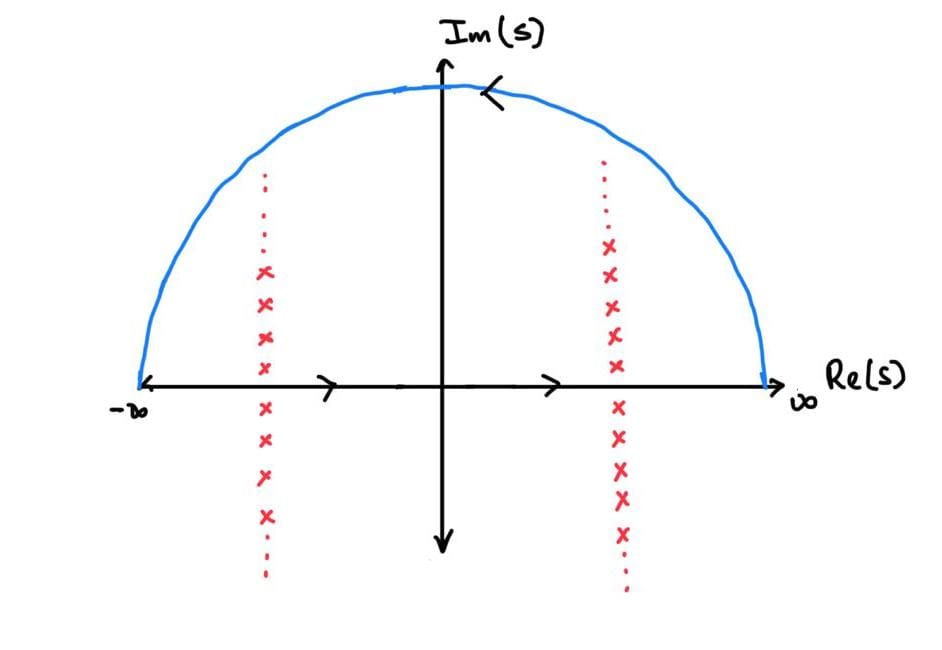}
\label{fig:upper}
}
\hfill
\subfloat[]{
\includegraphics[width=0.45\textwidth]{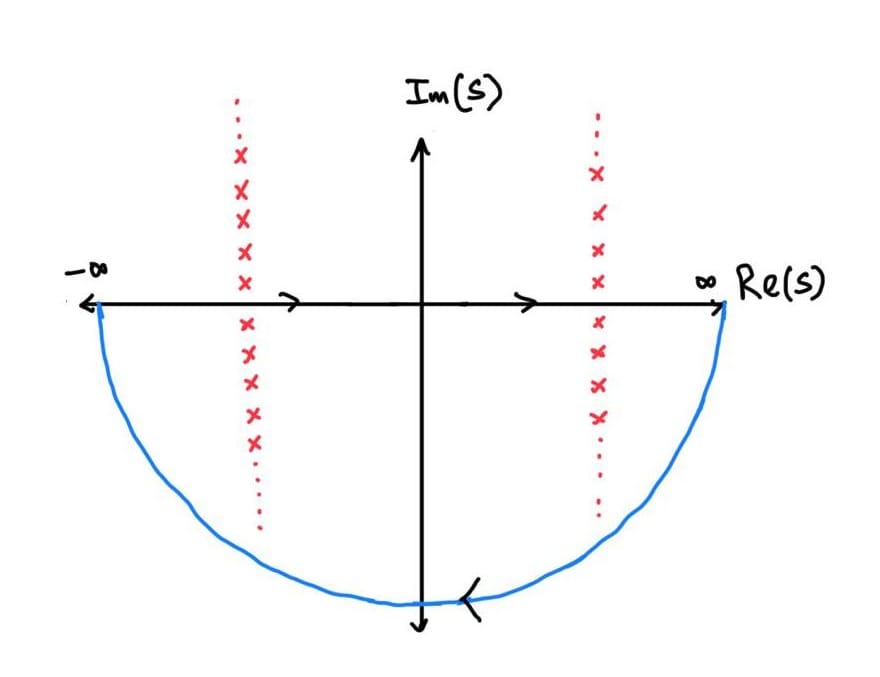}
\label{fig:lower}
}
\caption{(a) Contour of the integral in Eq.~(\ref{Ic}) for $\alpha>0$ and (b) contour of the integral in Eq.~(\ref{Ic}) for $\alpha<0$. The red dots represent the poles of the integrand.}
\label{fig:Polesfig}
\end{figure}
The contour integral is evaluated using the residue theorem. Since $\alpha>0$, we close the contour in the upper half of the complex $S$-plane. Therefore we have
\begin{equation}
\oint_C
\frac{e^{i\alpha S}}
{\sinh^2(S-i\epsilon)-K_{n1}^2}\,dS
=
2\pi i
\sum_{m=0}^{\infty}
\left[
\operatorname{Res}(S_1)
+
\operatorname{Res}(S_2)
\right].
\label{Ic1}
\end{equation}
The residue at the pole $S_1=u_{n1}+i\pi m+i\epsilon$ is
\begin{equation}
\begin{aligned}
\operatorname{Res}(S_1)
&=
\lim_{S\rightarrow S_1}
\frac{(S-S_1)e^{i\alpha S}}
{\sinh^2(S-i\epsilon)-K_{n1}^2} 
= \frac{e^{i\alpha S_1}}
{2\sinh(S_1-i\epsilon)\cosh(S_1-i\epsilon)}
\\
&=
\frac{
e^{i\alpha(u_{n1}+i\pi m+i\epsilon)}
}
{
2\left|K_{n1}\right|\sqrt{1+K_{n1}^2}
}.
\end{aligned}
\end{equation}
Similarly, the residue at the pole
$S_2=-u_{n1}+i\pi m+i\epsilon$ is
\begin{equation}
\begin{aligned}
\operatorname{Res}(S_2)
&=
\lim_{S\rightarrow S_2}
\frac{(S-S_2)e^{i\alpha S}}
{\sinh^2(S-i\epsilon)-K_{n1}^2}
= \frac{e^{i\alpha S_2}}
{2\sinh(S_2-i\epsilon)\cosh(S_2-i\epsilon)}
\\
&=
-
\frac{
e^{i\alpha(-u_{n1}+i\pi m+i\epsilon)}
}
{
2\left|K_{n1}\right|\sqrt{1+K_{n1}^2}
}.
\end{aligned}
\end{equation}
Substituting these in (\ref{Ic1}) and then using the Jordan lemma one finds
\begin{equation}
\begin{aligned}
I &=
-2\pi
\frac{\sin(\alpha  u_{n1})}
{\left|K_{n1}\right|\sqrt{1+K_{n1}^2}}
\sum_{m=0}^{\infty}
e^{-\alpha\pi m}.
\\
&=
-\frac{2\pi\sin(\alpha  u_{n1})}
{\left|K_{n1}\right|\sqrt{1+K_{n1}^2}}
\,
\frac{e^{\alpha\pi}}{e^{\alpha\pi}-1}.
\end{aligned}
\end{equation}
This leads to $\gamma_1(\omega_0)$ as,
\begin{equation}
    \gamma_1(\omega_0) = \frac{\hbar a}{4 \pi c^4}\sum_{n=-\infty}^{\infty}\frac{\sin(\alpha  u_{n1})}{\left|K_{n1}\right|\sqrt{1+K_{n1}^2}}
\,
\frac{e^{\alpha\pi}}{e^{\alpha\pi}-1}.
\end{equation}

Similarly, the second part corresponding to the second term of (\ref{C1}) can be evaluated . This yields
\begin{equation}
    \gamma_2(\omega_0) = \frac{\hbar a}{4 \pi c^4}\sum_{n=-\infty}^{\infty}\frac{\sin(\alpha  u_{n2})}{\left|K_{n2}\right|\sqrt{1+K_{n2}^2}}
\,\frac{e^{\alpha\pi}}{e^{\alpha\pi}-1}.
\end{equation}
Hence $\gamma(\omega_0)$ comes out to be as
\begin{equation}
    \gamma(\omega_0) = \gamma_1(\omega_0) - \gamma_2(\omega_0) = \frac{\hbar a}{4 \pi c^4}\frac{e^{\alpha\pi}}{e^{\alpha\pi}-1} \sum_{n=-\infty}^{\infty}\Bigg[\frac{\sin(\alpha  u_{n1})}{\left|K_{n1}\right|\sqrt{1+K_{n1}^2}}- \frac{\sin(\alpha  u_{n2})}{\left|K_{n2}\right|\sqrt{1+K_{n2}^2}} \Bigg]~.
\end{equation}
Now, we proceed to evaluate $\gamma(-\omega_0)$. In this case one has $\alpha<0$, for which the contour is chosen in the lower half of the complex $S$-plane, as shown in Fig.~\ref{fig:lower}. In this case, the enclosed poles correspond to $m=-1,-2,-3,\cdots$. Following the previous steps we obtain 
\begin{equation}
    \gamma(-\omega_0)= \frac{\hbar a}{4 \pi c^4}\frac{1}{e^{\alpha\pi}-1} \sum_{n=-\infty}^{\infty}\Bigg[\frac{\sin(\alpha  u_{n1})}{\left|K_{n1}\right|\sqrt{1+K_{n1}^2}}- \frac{\sin(\alpha  u_{n2})}{\left|K_{n2}\right|\sqrt{1+K_{n2}^2}} \Bigg]~.
\end{equation}
Finally, substituting these in (\ref{eq4}) and reintroducing the constant $\hbar$, we find the required expressions given in Eq. (\ref{eq21}). 

\section{Justification of the choice of max $n$} \label{App3}
In Section~\ref{Refl}, for the double-boundary case, we stated that the infinite summation can be safely truncated between $n= - 10^{8}$ to $n=10^8$ for the numerical calculations. To justify this choice, we examine the contribution of the summation term $J(nL)-J(nL+z_0)$. If this contribution becomes negligibly small beyond $|n|=10^{8}$, the terms beyond this limit will have a negligible effect on the summation, and the chosen truncation can be considered sufficient. However, since $J$ also depends on the acceleration $a$, the required truncation limit may vary with acceleration. Therefore, we specifically examine the case of $a=10^{5},\mathrm{m/s^2}$, which is the lowest acceleration considered in our analysis. If $J(nL)-J(nL+z_0)$ is already negligibly small at $n=10^{8}$ for this acceleration, then the chosen truncation limit is sufficient for our numerical analysis.

In Fig. \ref{fig:maxn} we plot  $J(nL)-J(nL+z_0)$ and a function of different choices of cutoff $n$ or $n_{max}$. 
\begin{figure}[!htbp]
\centering
\includegraphics[width=0.5\textwidth]{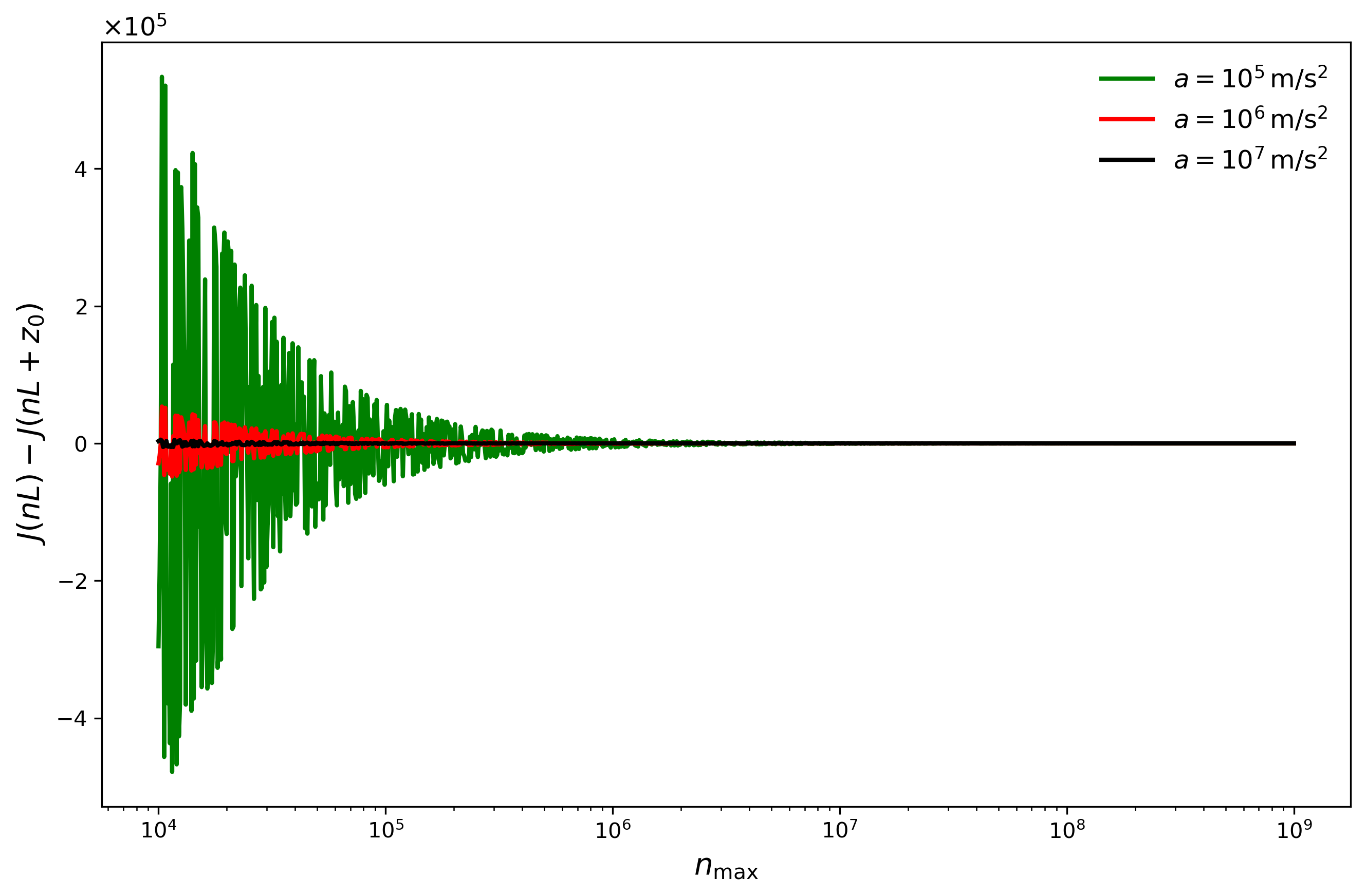}
\caption{Variation of $J(nL)-J(nL+z_0)$ as a function of the maximum summation index $n_{\max}$ for three different values of the acceleration, shown in different colors. The remaining parameters are fixed at $g=10^{-4}$ and $\omega_0=3\times10^9\,\mathrm{Hz}$.}
\label{fig:maxn}
\end{figure}
If the contribution $J(nL)-J(nL+z_0)$ becomes negligibly small beyond the chosen cutoff, the truncation can be considered sufficient. Figure~\ref{fig:maxn} shows that, for the lowest acceleration considered in our analysis, $a=10^5\,\mathrm{m/s^2}$, the value of $J(nL)-J(nL+z_0)$ becomes almost negligible well before $n_{max}=10^8$. Therefore, the contribution of the terms beyond $n_{max}=10^8$ is negligible, confirming that the chosen truncation is sufficient for our numerical analysis and does not significantly affect our results.

\section{Larger time plot}\label{App5}
In the previous sections, all the numerical results were obtained by taking the evolution time as $\tau=10^{-9}\,\mathrm{s}$. However, measuring or controlling such a short evolution time is experimentally challenging. Therefore, we also present the same analysis for a longer evolution time, namely $\tau=1\,\mathrm{s}$.

We first consider the thermal bath case. Here, we choose the background temperature to be $T=300\,\mathrm{K}$ and set the evolution time to $\tau=1\,\mathrm{s}$, while keeping all the other parameters unchanged. As shown in Fig.~\ref{fig:thermal1s}, both the behavior of the curve and the value of $\tau_{\mathrm{QSLT}}/\tau$ remain essentially unchanged compared with the case of $\tau=10^{-9}\,\mathrm{s}$.
\begin{figure}[!htbp]
\centering
\includegraphics[width=0.4\textwidth]{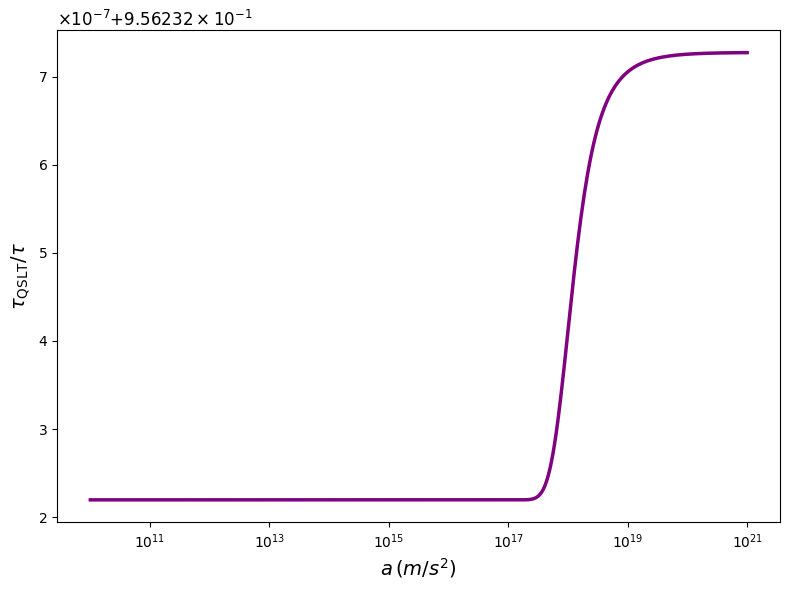}
\caption{Variation of the normalized quantum speed limit time, $\tau_{\mathrm{QSLT}}/\tau$, as a function of acceleration for background temperatures $300~\mathrm{K}$. The remaining parameters are fixed at $\theta=\pi/4$, $\tau=1\,\mathrm{s}$, and $g=10^{-5}$.}
\label{fig:thermal1s}
\end{figure}
Now, we consider the single-boundary case with the evolution time $\tau=1\,\mathrm{s}$. Although the overall behavior of the curve remains unchanged, the value of $\tau_{\mathrm{QSLT}}/\tau$ is different from that obtained for $\tau=10^{-9}\,\mathrm{s}$. However, this change occurs only up to $\tau=10^{-7}\,\mathrm{s}$. For evolution times greater than $10^{-7}\,\mathrm{s}$, $\tau_{\mathrm{QSLT}}/\tau$ becomes independent of the evolution time. To demonstrate this, we present two plots, one for $\tau=1\,\mathrm{s}$ and another for a much larger evolution time, $\tau=10^{5}\,\mathrm{s}$. As shown in Fig.~\ref{fig:single1s1} and \ref{fig:single1s1} , the two curves completely overlap, confirming that no further change occurs for $\tau \geq 10^{-7}\,\mathrm{s}$.
\begin{figure}[!h]
\centering
\subfloat[]{
\includegraphics[width=0.45\textwidth]{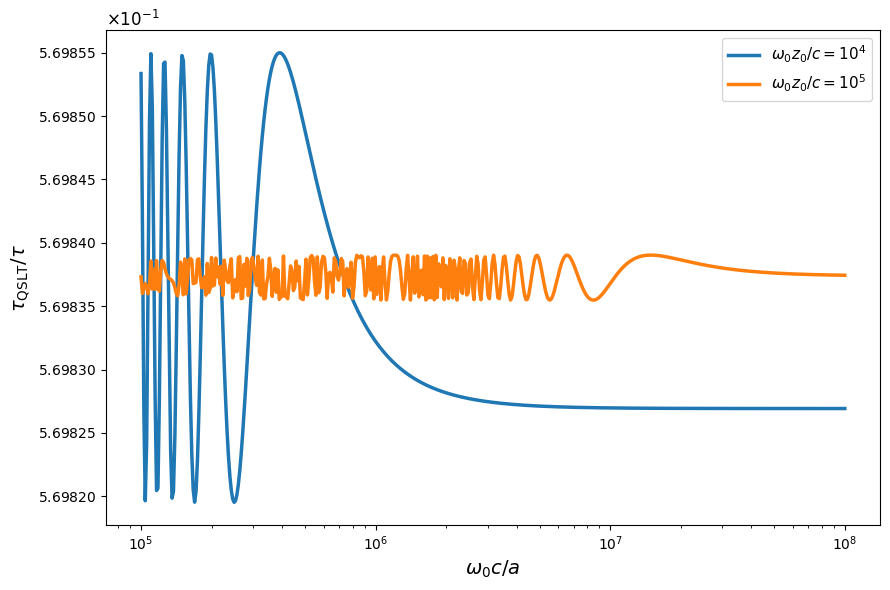}
\label{fig:single1s1}
}
\hfill
\subfloat[]{
\includegraphics[width=0.45\textwidth]{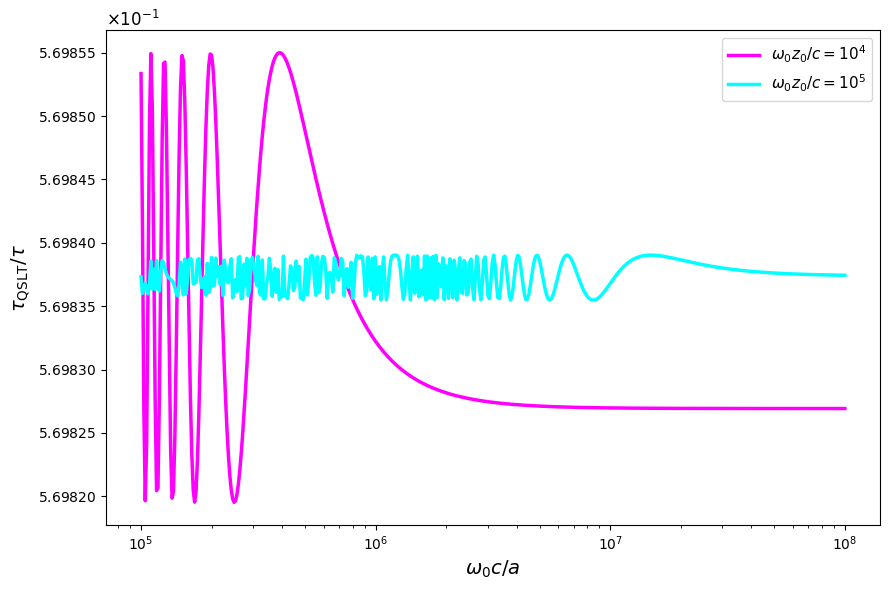}
\label{fig:single1s1}
}
\caption{Variation of the normalized quantum speed limit time, $\tau_{\mathrm{QSLT}}/\tau$, as a function of the acceleration $\omega_0 c/a$ ($\mathrm{m/s^2}$) in present of single boundary, for two different values of the dimensionless parameter $\omega_0 z_0/c = 10^4$ and $10^5$ and for two different evolution time: (a)$\tau=1\,\mathrm{s}$, (b)$\tau=10^{5}\,\mathrm{s}$. The remaining parameters are fixed at $\theta=\pi/4$, $g=10^{-4}$, and $\omega_0 = 3 \times 10^9 \mathrm{Hz}$.}
\label{fig:singlelargetime}
\end{figure}
Next, we investigate the normalized quantum speed limit time, $\tau_{\mathrm{QSLT}}/\tau$, for the double-boundary case with a longer evolution time, $\tau=1\mathrm{s}$. This allows us to examine whether the behavior observed for the shorter evolution time remains unchanged during a longer-time evolution.
\begin{figure}[!htbp]
\centering
\includegraphics[width=0.5\textwidth]{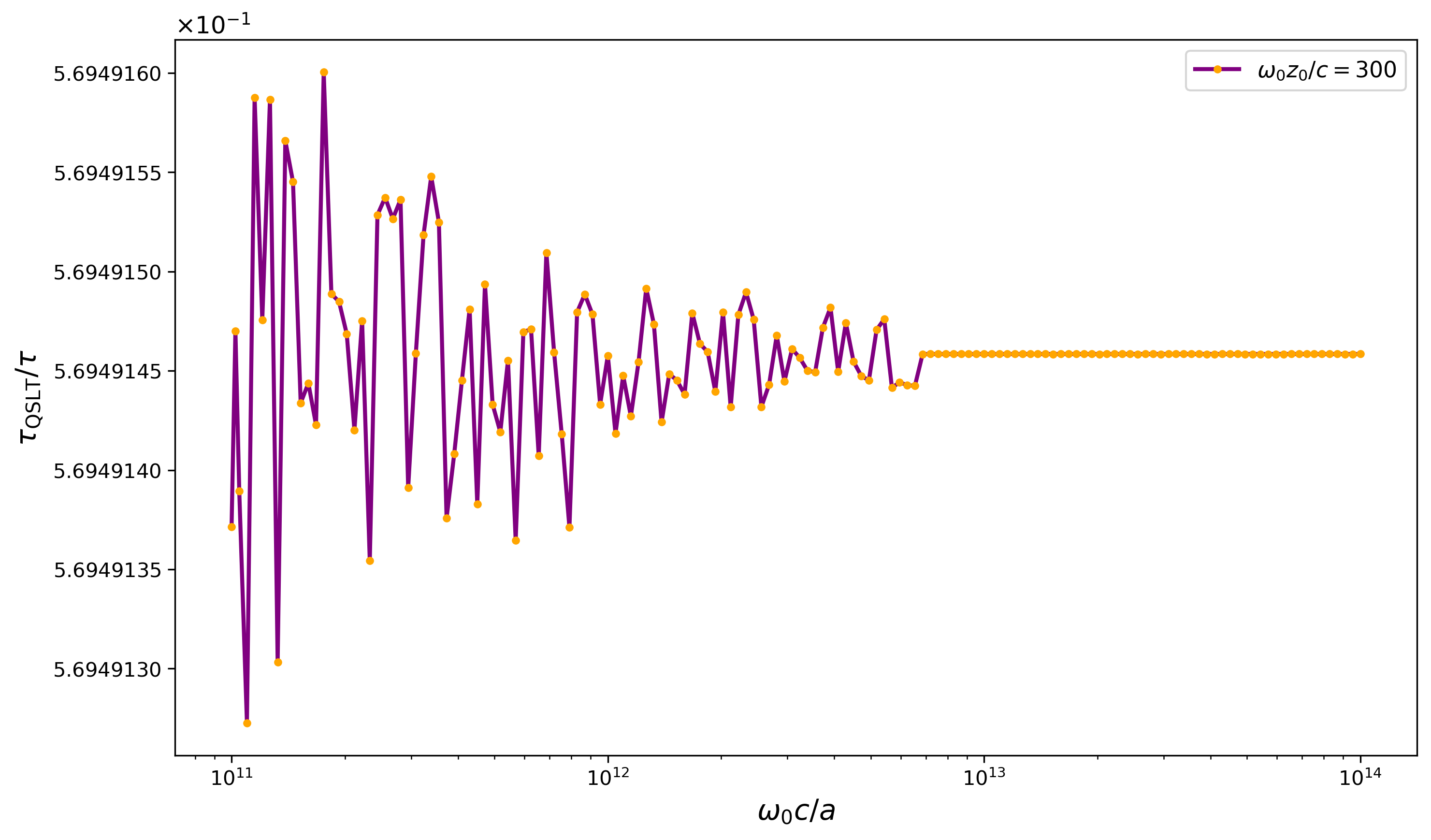}
\caption{Variation of the normalized quantum speed limit time, $\tau_{\mathrm{QSLT}}/\tau$, as a function of the dimensionless acceleration parameter $\omega_0 c/a$ for the double-boundary configuration, with $\omega_0 z_0/c=300$ and $\omega_0 L/c=3000$. The evolution time is $\tau=1\,\mathrm{s}$.The yellow dots represent the values obtained from Eq.~(\ref{eq10}), while the purple line connects these points. The remaining parameters are fixed at $\theta=\pi/4$, $g=10^{-4}$, and $\omega_0=3\times10^9\,\mathrm{Hz}$.}
\label{fig:double1s}
\end{figure}

Here, our main objective is to identify the footprint of the Unruh effect. Although the value of $\tau_{\mathrm{QSLT}}/\tau$ decreases, the reduced value remains experimentally accessible. Moreover, the overall shape of the curve remains unchanged. Therefore, this reduction in $\tau_{\mathrm{QSLT}}/\tau$ is not expected to significantly affect the detection of the Unruh effect.

\end{widetext}

\makeatletter
\makeatother
\bibliographystyle{unsrt}
\bibliography{biblio}

\end{document}